\documentclass[10pt,aps,prb,twocolumn,amsmath,amssymb,
longbibliography,
superscriptaddress
]{revtex4-2}
\usepackage{graphicx}
\usepackage{epsfig}
\usepackage{dcolumn}
\usepackage{bm}
\usepackage{bbm}
\usepackage{array}
\usepackage[normalem]{ulem}
\usepackage{color}
\usepackage{float}
\usepackage[colorlinks=true,linkcolor=blue,citecolor=blue,urlcolor=blue,breaklinks=true]{hyperref}
\usepackage{verbatim}
\usepackage{subfigure}
\usepackage[ruled,vlined]{algorithm2e}
\usepackage{tikz}
\usepackage{braket}

\def\be{\begin{equation}}       \def\ee{\end{equation}}
\def\bea{\begin{eqnarray}}      \def\eea{\end{eqnarray}}
\def\ba{\begin{array}}
\def\ea{\end{array}}
\def\bnum{\begin{enumerate} }
\def\enum{\end{enumerate}}

\def\=>{\Rightarrow}
\def\>{\rightarrow}

\def\eye2{Fathbb{I}}

\usepackage{ listings }

\def\Eq#1{Eq.~(\ref{#1})}
\def\Fig#1{Fig.~\ref{#1}}

\renewcommand{\>}{\rangle}

\newcommand{\eq}[2]{
	\begin{equation}
	#1 \label{#2}
	\end{equation}
}

\newcommand{\mi}{\mathrm{i}}

\renewcommand{\rm}[1]{\mathrm{#1}}

\newcommand{\vect}[1]{\boldsymbol{#1}}

\usepackage{listings}
\usepackage{color}
\definecolor{lightgray}{gray}{1}

\newtheorem{inequality}{Inequality}

\newcommand\COMMENTED[1] {}

\newcommand\redsout{\bgroup\markoverwith{\textcolor{red}{\rule[0.5ex]{2pt}{0.4pt}}}\ULon}
\begin{document}

\title{Revisiting Nishimori multicriticality through the lens of information measures}

\author{Zhou-Quan Wan}
\thanks{The two authors contributed equally to this work.}
\affiliation{Center for Computational Quantum Physics, Flatiron Institute, New York, NY 10010, USA}

\author{Xu-Dong Dai}
\thanks{The two authors contributed equally to this work.}
\affiliation{New Cornerstone Science Laboratory, Department of Physics,\\ \text{The Hong Kong University of Science and Technology,
Clear Water Bay, Kowloon 999077, Hong Kong, China}}

\author{Guo-Yi Zhu}
\email{guoyizhu@hkust-gz.edu.cn}
\affiliation{\text{The Hong Kong University of Science and Technology (Guangzhou), Nansha, Guangzhou, 511400, Guangdong, China}}

\begin{abstract}
The quantum error correction threshold is closely related to the Nishimori physics of random statistical models.
We extend quantum information measures such as coherent information beyond the Nishimori line and establish them as sharp indicators of phase transitions over the full $p$-$T$ plane.
These generalized measures admit a natural operational interpretation as diagnostics of inference mismatch for decoders operating at an effective temperature.
We derive exact inequalities for several generalized measures, demonstrating that each attains its extremum along the Nishimori line.
As a direct application, we study these measures in the 2d $\pm J$ random-bond Ising model—corresponding to a surface code under bit-flip noise—and revisit the Nishimori multicritical point.
Among all indicators, coherent information exhibits the weakest finite-size effects, enabling a high-precision estimate $p_c=0.1092212(4)$ and the associated critical exponents.
\end{abstract}

\date{\today}
\maketitle

\textit{Introduction.---} 
The random-bond Ising model (RBIM) is one of the foundational models of disordered systems and has attracted sustained interest for decades in statistical mechanics.
Owing to its gauge invariance, a special manifold in the parameter space—known as the Nishimori line—emerges.
Along this line, exact results—such as the internal energy and constraints on the phase diagram—have been obtained
\cite{Nishimori1981,Georges1985exactII,iba1999nishimori,Nishimori2001book}.
Remarkably, in quantum error correction (QEC), the error threshold maps directly onto the Nishimori physics of the corresponding statistical model \cite{TQM,WANG200331,KatzgraberPRL2009,BombinPRA2010,BombinPRX2012,RevModPhys.87.307,Chubb_2021,Benjamin2023CoherentError}, with the optimal threshold corresponding to the multicritical Nishimori point (MNP)~\cite{Bravyi2014,Bonilla2021xzzx,PhysRevLett.120.050505,Bravyi2019BiasedNoise,Xiao2024exactresultsfinite,Zhao2024ApproxQEC}.
Recently, the coherent information—which quantifies how much quantum information can be reliably transmitted through a noisy channel~\cite{Nielsen96coherentinfo,Lloyd97coherentinfo,schumacher2001approximatequantumerrorcorrection}—has also been shown to exhibit a deep connection to the RBIM \cite{Fan2024mixedTO,cenke2023prxq,Wang2025MixedTO,LeePRL2025,LeePRA2025,huang2024coherentinfo}. Notably, it shows remarkably small finite-size effects in studying the information-theoretic transition~\cite{MarkusPRR2024,GY2024prxq,huang2024coherentinfo,colmenarez2025fundamentalthresholdscomputationalerasure,Wang25selfdual,gy2025learningtransition}.
The Nishimori physics has also gained renewed significance through its connections to monitored quantum circuit dynamics \cite{GY2023prl,chen2025nishimori,gy2025nishimoriuniversality}, mixed-state phase transitions \cite{Fan2024mixedTO,cenke2023prxq,Wang2025MixedTO,Perm2025MixedTO,Ellison2025MixedTO,yyzprb2024,chong2025prxq,liu2025diagnosingswssb,Tarun2024Selfdual,PhysRevB.110.085158} and statistical inference problems \cite{Zdeborova16inferencethreshold, gy2025learningtransition, Nahum25bayesiancriticalpoints, Lamacraft25inference, Vasseur25fluctuatinghydrodynamics}.

In the classical phase diagram of RBIM, the MNP exhibits a distinctive multicritical nature: both the Nishimori line and temperature act as independent, relevant perturbations \cite{Doussal1988locationMNP,Harris1989prbepsilonexpansion}. Despite this, most existing studies have focused either strictly on the Nishimori line or on the zero-temperature limit, which corresponds to the optimal or the matching-type decoder. In contrast, the impact of temperature deformations away from the MNP—and their implications for inference and quantum error correction—remains largely unexplored.
Moreover, previous studies on RBIM~\cite{DWRG,nishimori2002duality,Ohzeki2015,nishimori2007duality,Nishimori1987MCRG,Ettore2008MC,Ettore2009strong,nishimori1987TM,ReisPRB1999TM,HoneckerPRL2001,Picco2006CentralCharge,LessaPRB2006TM,TM2009prb,Cho1997PRB,Chalker2002Mapping,Chalker2002NegativeScaling,Haijun2014TN,tebd2022,YoujinPRB2025} have generally been limited to small system sizes and have rarely given a systematic comparison between statistical-mechanical observables and quantum information–theoretic quantities.

Here we develop a unified framework that extends information measures—most notably coherent information $\mathcal{I}_c(\beta)$—beyond the Nishimori line ($\beta_p$) to arbitrary temperature ($\beta_p$) over the full $p$–$T$ plane. 
In this statistical formulation, $\mathcal{I}_c(\beta)$ (defined by the disorder-averaged log-posterior) acquires a direct operational meaning as decoding with an effective temperature $\beta$: it quantifies the mismatch between the true error distribution and the inference distribution produced by the Bayes decoder. 
All generalized measures can be expressed in terms of the domain-wall free energy (DWFE) $\Delta F$, a standard probe of phase transitions via its disorder average and general moments \cite{DWRG,Chalker2002NegativeScaling}, thereby turning them into sharp statistical indicators away from the Nishimori line. 
Finally, exploiting gauge invariance, we derive inequalities showing that $\mathcal{I}_c(\beta)$ and related observables attain their extrema on the Nishimori line. 
These exact constraints clarify the performance of mismatched decoders, single out the optimal decoder, and impose nontrivial structure on the phase diagram.

With these new insights and improved estimators based on the Nishimori relation, we perform large-scale simulations using the fermionic transfer-matrix method~\cite{Chalker1988,Cho1997PRB,Read2000,Gruzberg2001,Chalker2002Mapping,Chalker2002NegativeScaling,Bravyi2014,ChaomingPRB2022,Bravyi2005FGO,Li2016PRL,han2024pfaffian} and obtain mutually consistent critical-point estimates, $p_c$=$0.1092212(4)$, representing, to our knowledge, the most precise value reported to date.
Along the Nishimori line, the coherent information exhibits exceptionally small finite-size effects, whereas temperature deviations from the MNP reveal that the domain-wall entropy shows similarly minimal finite-size corrections. 
We further verify that the DWFE distribution is scale-invariant at the MNP~\cite{Chalker2002NegativeScaling}, explaining the coincidence between the MNP and the QEC threshold.

\begin{figure*}[t]
    \centering
    \includegraphics[width=1.0\linewidth]{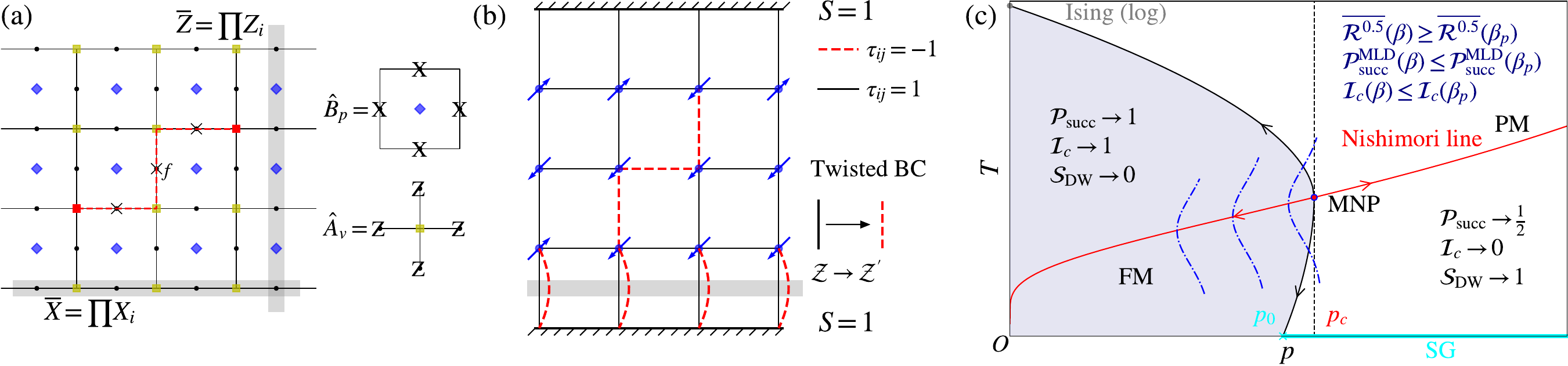}
    \caption{\textbf{Surface code, RBIM and schematic phase diagram.} (a) Surface code with rough (smooth) boundary conditions along the $x$ ($y$) direction, for code distance $d=4$. 
    Blue (yellow) markers indicate plaquette (vertex) stabilizers;
    The shaded regions indicate qubits defining logical $\bar X$ ($\bar Z$) operators. 
    The dashed red line shows an error chain $f$, with its endpoint marking the syndrome $\mathbf{s}$.
    (b) RBIM ($L=4$) mapped from panel (a), with spins on plaquettes operators and qubits are mapped to the bonds between them.
    The rough (smooth) boundary condition corresponds to free (fixed) boundary conditions in the RBIM, and physical errors appear as $\tau_{ij}$=$-1$ bonds (red dashed lines).
    Inserting a domain wall (logical error $\bar{X}$) corresponds to imposing a twisted boundary condition.
    (c) Schematic RBIM phase diagram \cite{Ettore2008MC} in the $p$-$T$ plane, showing paramagnetic (PM), ferromagnetic (FM) and zero-temperature spin glass (SG) phases. The Nishimori line (red) represents the mapping from panel (a) to (b) in \Eq{eq:mapping}; the thermodynamic-limit behavior of various measures in each phase is indicated. The MNP marks the intersection of the FM–PM phase boundary with the Nishimori line and features two relevant directions (arrows) in the RG sense.
    Dashed blue curves schematically illustrate the extreme behavior of $\overline{\mathcal{R}^{0.5}(\beta)},\mathcal{P}_\text{succ}^\text{MLD},\mathcal{I}_c(\beta)$ along the Nishimori line.
    }
    \label{fig:surface_mapping}
\end{figure*}

\textit{Surface code and RBIM.---}
We begin by outlining the mapping from the surface code to the RBIM.
As shown in \Fig{fig:surface_mapping}(a), the surface code is defined on a square lattice with qubits placed on edges. 
Its Hamiltonian reads $\hat H = - \sum_{v} \hat {A}_v- \sum_p \hat {B}_p$ \cite{KITAEV20032},
where the plaquette operator $\hat B_p = \prod_{i \in \partial p} X_i$ and the vertex operator $\hat A_v = \prod_{i \in v} Z_i$ are graphically represented in Fig.~\ref{fig:surface_mapping}(a).
All $\hat B_p$ and $\hat A_v$ operators commute with each other, so the ground state is stabilized by these operators.
In the lattice geometry of \Fig{fig:surface_mapping}(a), the number of independent stabilizers is one fewer than the number of qubits. As a result, the ground state is two-fold degenerate, allowing the surface code to encode a single logical qubit \cite{Bravyi2014}.
The logical $\bar Z$($\bar X$) operator is defined as the product of $Z$($X$) operators along the vertical (horizontal) direction, as shown in \Fig{fig:surface_mapping}(a).

Now, we consider an error channel $\mathcal{N}=\otimes_i\mathcal{N}_i$ acting on a quantum state $\rho$ with $\mathcal{N}_i[\rho] =(1-p)\rho+pX_i\rho X_i$, where $p$ is the bit-flip error rate.
As shown in \Fig{fig:surface_mapping}(a), a bit-flip error chain $f$ with $\prod_{l\in f}X_l$,
creates $e$-anyon excitations at its endpoints defined by $\hat{A}_v$ =$ -1$~\cite{KITAEV20032}.
These excitations, commonly referred to as the error syndrome and denoted by $\mathbf{s}$.
All error configurations $f'$ satisfying $\mathbf{s}=\partial f'$ (i.e., exhibiting the same observable syndromes detected by $\hat{A}_v$ for every $v$) fall into two equivalence classes: the homology-trivial class $[f]$, in which the elements differ from $f$ by local $\hat{B}_p$ operations, and the homology-nontrivial class $[f\bar X]$, generated by the logical operator $\bar X$ (Fig.~\ref{fig:surface_mapping}(a)).
QEC aims to identify which class the actual error belongs to; recovery succeeds if the correction and true error are in the same class, and fails otherwise~\cite{TQM,RevModPhys.87.307}.
Consequently, it is crucial to evaluate the probability that an error configuration belongs to the class $[f]$.
It has been shown that \cite{TQM,Bravyi2014}
\eq{
    P([f])\equiv \sum_{f'\in [f]}P(f')={\left(2\cosh(\beta_p)\right)^{-N}}\mathcal{Z}(\{\tau_{ij}\},\beta_p),
}{eq:mapping}
where $\mathcal{Z}(\{\tau_{ij}\},\beta) = \sum_{S_i=\pm1} \exp(\beta\tau_{ij}S_i S_j)$ is the partition function of the RBIM (see \Fig{fig:surface_mapping}(b)) with $\tau_{ij} $=$-1$ for bond $ij$$\in$$f$ ($1$ otherwise), the inverse temperature $ \beta_p = -\frac 1 2 \log{\frac{p}{1-p}}$ defines the Nishimori line, and $N$ the number of qubits.
The probability of the other class, $P([f\bar X])$, corresponds to the same model with \emph{twisted boundary conditions}, obtained by inserting a non-contractible domain wall that exchanges ferromagnetic and antiferromagnetic couplings, denoted by $\mathcal Z'(\{\tau_{ij}\},\beta_p)$.

\textit{Information Measures and Statistical Indicators.---}
Several strategies exist for determining the error class and performing correction.
One is the maximum-likelihood decoder (MLD), which compares $P([f])$ and $P([f\bar X])$, selecting the more probable class \cite{RevModPhys.87.307}.
The success probability of this decoder can then be interpreted as the disorder-averaged probability that the correct class is favored.
Using \Eq{eq:mapping}, it can be written as
\eq{
\mathcal P_\text{succ}^\text{MLD}(\beta_p) = \overline{\Theta(\mathcal Z(\{\tau_{ij}\},\beta_p)-\mathcal Z'(\{\tau_{ij}\},\beta_p))},
}{eq:psucc}
where $\Theta (x)$ is the Heaviside step function ($\Theta(0)=\frac 1 2 $), and $\overline{\cdot}$ denotes the disorder average over random configurations $\{\tau_{ij}\}$.
By construction, the MLD maximizes the success probability among all decoders.
We can also consider a Bayes-optimal decoder that outputs the posterior distribution over logical classes, $P([f]|\mathbf{s}) = \frac{P([f])}{P([f])+P([f\bar X])}$, which has a success probability
\eq{
\mathcal P_\text{succ}^\text{Bayes}(\beta_p) = \overline{\frac{\mathcal Z(\{\tau_{ij}\},\beta_p)}{\mathcal Z(\{\tau_{ij}\},\beta_p)+\mathcal Z'(\{\tau_{ij}\},\beta_p)}}.}{eq:log_likelihood}
The Bayes decoder is also optimal in the sense that it maximizes the log-posterior $\overline{\log_2 P([f]|\mathbf{s})}$.
It is closely related to the coherent information~\cite{Nielsen96coherentinfo,Lloyd97coherentinfo,schumacher2001approximatequantumerrorcorrection}, which is an information-theoretic diagnostic quantifying the logical information surviving the error channel, independent of specific decoding algorithms.
For surface code under bit-flip error, the logical coherent information was recently derived~\cite{Fan2024mixedTO,cenke2023prxq,Wang2025MixedTO}, relating to the averaged log-posterior:
\eq{
\mathcal I_c (\beta_p) = 1+\overline{\log_2\left(\frac{\mathcal Z(\{\tau_{ij}\},\beta_p)}{\mathcal Z(\{\tau_{ij}\},\beta_p)+\mathcal Z'(\{\tau_{ij}\},\beta_p)}\right)}.
}{eq:coherent_info}
Together, Eqs.~\eqref{eq:psucc}–\eqref{eq:coherent_info} establish a mapping between quantum information–theoretic quantities and statistical variables of the RBIM along the Nishimori line. The mapping extends beyond it by replacing $\beta_p$ with a general $\beta$, interpretable as an \emph{effective inverse temperature} for decoders with $\beta \neq \beta_p$, as occurs when the physical error rate $p$ is unknown~\cite{Zdeborova16inferencethreshold, DanielPRA2014}. For instance, the minimum-weight perfect matching decoder, independent of $p$, corresponds to the zero-temperature limit $\beta\to\infty$~\cite{TQM}. For general $\beta\neq\beta_p$, $\mathcal{I}_c(\beta)$ no longer coincides exactly with the coherent information. Nevertheless, from an error-correction perspective, $\mathcal{I}_{c}(\beta)$ quantifies the mismatch between the actual error distribution and the inference distribution produced by the Bayes decoder. 

Recently, the domain-wall entropy was proposed to capture phase transitions in both quantum information and classical inference \cite{gy2025learningtransition,gy2025nishimoriuniversality}.  
Consider associating the domain wall with an auxiliary spin that takes the value $-1$ ($+1$) if a domain wall is (is not) inserted. 
This allows us to define the entropy of this spin as
\begin{equation}
    \mathcal{S}_\text{DW}(\beta) = \overline{ - 
    \frac{\mathcal{Z}}{\mathcal{Z}+\mathcal{Z}'} 
    \log_2 \frac{\mathcal{Z}}{\mathcal{Z}+\mathcal{Z}'} 
    - (\mathcal{Z} \leftrightarrow \mathcal{Z}') }.\label{eq:dwentropy}
\end{equation}
Along the Nishimori line, $\mathcal I_c(\beta_p) = 1-\mathcal{S}_\text{DW}(\beta_p)$~\cite{GY2024prxq}, making $\mathcal{S}_\text{DW}(\beta)$ an alternative generalization of coherent information.
However, as shown later, $\mathcal{S}_\text{DW}(\beta)$ and $\mathcal{I}_c(\beta)$ exhibit markedly different temperature dependencies.

From a statistical perspective, the limiting behavior of these quantities in different phases can be understood directly from the domain-wall free energy (DWFE) cost
\begin{equation}
    \Delta F(\{\tau_{ij}\},\beta) \equiv \log\mathcal Z(\{\tau_{ij}\},\beta)-\log \mathcal Z'(\{\tau_{ij}\},\beta),
\end{equation}
conditioned upon the error configuration.
All quantities in Eqs.~\eqref{eq:psucc}–\eqref{eq:dwentropy} are determined by the distribution of $\Delta F$.
In the ferromagnetic (FM) phase, the average $\Delta F$ grows linearly with system size, making domain-wall flips exponentially unlikely; thus, $\mathcal{P}_{\text{succ}},\mathcal{I}_{c}\to 1$, indicating the success of error correction and no loss of information.
In the paramagnetic (PM) phase, the domain wall loses its rigidity and $\Delta F \to 0$, yielding $\mathcal{P}_{\text{succ}} \to 1/2$ and $\mathcal{I}_{c}$ goes to 0, signifying the failure of error correction and the loss of information.
Accordingly, the domain-wall entropy $\mathcal S_\text{DW}$ rises from 0 to 1 across the FM-PM transition.
In this broader context, $\mathcal{P}_{\text{succ}}$, $\mathcal{I}_c$ and $\mathcal S_\text{DW}$ not only retain their significance in quantum information but can also be interpreted as statistical indicators of the phase transition (See Fig. \ref{fig:surface_mapping}(c)).

The properties of the domain wall have long been used in statistical studies of the RBIM; in particular, the disorder-averaged DWFE, defined as 
\begin{equation}
    d_W(\beta)\equiv \overline{\log\mathcal Z(\{\tau_{ij}\},\beta)-\log \mathcal Z'(\{\tau_{ij}\},\beta)}=\overline{\Delta F},\label{eq:DWFE}
\end{equation}
is widely employed to characterize the phase transition \cite{DWRG,HoneckerPRL2001}.
As a disordered ensemble, one can consider its generic $q$-th moment average, defined as:
\eq{
\overline{ \mathcal{R}^q(\beta)}\equiv \overline{ \left(\mathcal Z'(\{\tau_{ij}\},\beta)/\mathcal Z(\{\tau_{ij}\},\beta)\right)^q} 
= \overline{e^{-q \cdot \Delta F}}.
}{eq:ratiopow}
This quantity is related to the disorder operator in quasi-one-dimensional geometries and has been used to study the multifractal spectrum of the MNP~\cite{Read2000,Chalker2002NegativeScaling}.  

Note that all these disorder-averaged observables are gauge-invariant, remaining unchanged under the transformation $\tau'_{ij}$=$\tau_{ij} \theta_i \theta_j$ for any $\theta_i$=$\pm 1$. 
Using this property, we can show that $\mathcal{P}_\text{succ}^\text{MLD}, \mathcal{I}_c, \overline{\mathcal{R}^{0.5}}$ attain their extremum at $\beta_p$
(see proofs in App.~\ref{app:end_matter}).
This reflects the optimality of the MLD decoder, which maximizes success probability, and the Bayes decoder, whose inference distribution best matches the true error distribution; $\mathcal I_c$ and $\overline{\mathcal{R}^{0.5}}$ quantify this distribution mismatch.
In Fig.~\ref{fig:surface_mapping}(c), we show a schematic RBIM phase diagram, summarizing the asymptotic behavior of these quantities across phases and the inequalities discussed above.

In the following, we present the numerical results for these quantities, demonstrating their effectiveness in locating the MNP and characterizing the surrounding critical behavior.
We also examine the $\Delta F$ distribution to support the discussion above.
Further algorithmic details and numerical stability analysis are given in App.~\ref{app:alg} with an open-source implementation~\footnote{\href{https://github.com/Zhouquan-Wan/fermionic-transfer-matrix-rbim}{https://github.com/Zhouquan-Wan/fermionic-transfer-matrix-rbim}}.

\begin{figure*}
    \centering
    \includegraphics[width=1.0\linewidth]{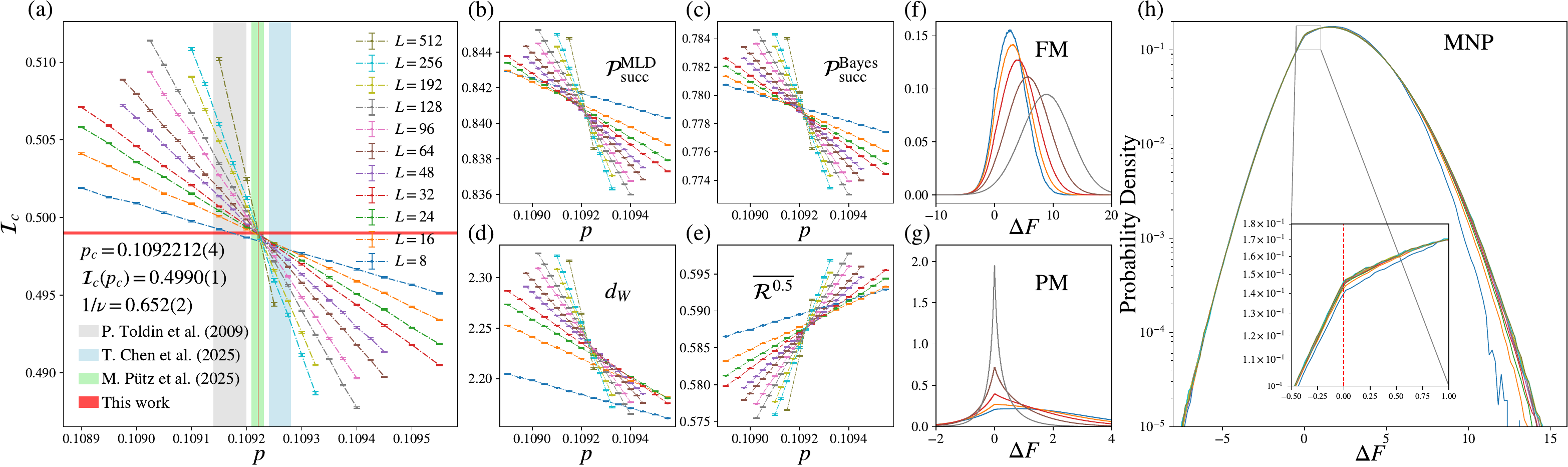}
    \caption{{\bf Critical behavior of various quantities along the Nishimori line}. Finite-size scaling yields $p_c = 0.1092212(4)$ and $1/\nu = 0.652(2)$.
    (a) Coherent information $\mathcal I_c$; the shaded region (red) indicates our critical-point result, compared with previous works~\cite{Ettore2009strong,YoujinPRB2025,gy2025nishimoriuniversality}.
    The horizontal region shows the crossing value $\mathcal{I}_c=0.4990(1)$, deviating clearly from 1/2.
    (b) The success probability of MLD $\mathcal P_\text{succ}^{\text{MLD}}$. 
    (c) The success probability of the Bayes decoder $\mathcal P_\text{succ}^{\text{Bayes}}$. (d) Disorder-averaged DWFE $d_W$. 
    (e) $q$=$0.5$ moment of partition function ratio $\overline{\mathcal R^{0.5}}$, which is a special case of \Eq{eq:ratiopow}. 
    (f),(g) Distribution of $\Delta F$ in the FM ($p$=$0.10$) and PM ($p$=$0.12$) phases.
    (h) The distribution of $\Delta F$ at MNP, exhibiting a characteristic kink at zero. 
    Quantities shown in (a–e) were computed with an estimator incorporating the Nishimori condition to reduce statistical errors, see \Eq{eq:better_estimator}; Separate fitting results are summarized in Table~\ref{tab:end_matter_res_tab}, and further details of the finite-size scaling analysis can be found in App.~\ref{app:fss}.
    }
    \label{fig:crossing}
\end{figure*}

\textit{Results along the Nishimori line.---}
In our numerical calculations, we consider a square geometry with open (fixed) boundary conditions along the \(x\)(\(y\))-direction, as shown in \Fig{fig:surface_mapping}(b).  
\Fig{fig:crossing}(a--e) show the quantities defined above along the Nishimori line. 
Using the Nishimori relation, we employ an improved estimator,
$\overline{O(\tau)} =\overline{\frac{\mathcal{Z}}{\mathcal{Z}+\mathcal{Z}'}O+\frac{\mathcal{Z'}}{\mathcal{Z}+\mathcal{Z}'}O'}$, which reduces statistical errors ($O'$ for twisted boundary condition).
With system size up to $512$ and over $10^7$ disorder configurations, 
we determine the critical point to be $p_c=0.1092212(4)$ and the correlation length exponent $1/\nu=0.652(2)$, improving the precision of $p_c$ by roughly two digits over previous most accurate estimates \cite{Ettore2009strong,YoujinPRB2025,gy2025nishimoriuniversality}.
The numerical value also lies clearly {\it below}, though close to, the duality-conjectured~\cite{nishimori2002duality} analytic value 0.1100... derived from the condition $-p\log_2 p - (1-p)\log_2 (1-p)=1/2$, also known as the Hashing bound~\cite{TQM, Bennett96hashingbound} in the context of quantum error correction. 
Our estimate of $1/\nu$ also shows substantial improvement over previous results on RBIM, including $ 0.66(1)$~\cite{Ettore2009strong} and $0.67(1)$~\cite{YoujinPRB2025}, and is in tension with a recent conjecture $1/\nu$=$2/3$~\cite{Delfino2024CriticalEA}.
At the same time, it agrees well with the recent calculation of the Nishimori universality class in monitored circuits, $\nu = 1.532(4)$~\cite{gy2025nishimoriuniversality}.
We extract the anomalous dimension by fitting the magnetic ordering, obtaining $\eta = 0.1786(6)$ (see App.~\ref{app:fss}). This value is consistent with previous estimates, $\eta = 0.177(2)$~\cite{Ettore2009strong} and $\eta = 0.180(1)$~\cite{YoujinPRB2025}.

Notably, the consistency of $p_c$ across observables within this narrow interval provides high-precision confirmation of the coincidence between the MNP and the QEC threshold.
As shown in \Fig{fig:crossing}, the coherent information exhibits the smallest finite-size corrections, with a crossing point $\mathcal I_c=0.4990(1)$ very close to $\frac 1 2$.
In Ref.~\cite{huang2024coherentinfo}, the coherent information $\mathcal{I}_c^{(n)}$ in $n$-replica systems was shown to be exactly $\frac 1 2$ at $n=2,3$ due to self-duality, leading to the conjecture that $\mathcal{I}_c$ also equals $\frac 1 2$ as $n\rightarrow 1$. 
A related duality method previously yielded the estimate $p_c = 0.10929(2)$ in Ref.~\cite{Ohzeki2015}.
Our results indicate that this self-duality does not hold exactly in this limit, since $\frac 1 2 $ lies well outside the error bar, yet it offers a valuable explanation for the remarkably small finite-size effects in $\mathcal{I}_c$
\footnote{In our lattice setup, self-duality holds exactly for replica $n = 2,3$ only at aspect ratio $L_x/L_y=1$. This is important because simulations with different aspect ratios with open and fixed boundary conditions can lead to significantly larger finite-size effects and cause the crossing point to deviate substantially from $1/2$.}.
It is consistent with the fact that in the long wavelength limit, the Nishimori criticality breaks Kramers-Wannier duality~\cite{Chalker2002NegativeScaling}, whose restoration results in a distinct critical theory~\cite{Gruzberg2001, Wang25selfdual}. 

As discussed earlier, all the quantities considered are related to the distribution of $\Delta F$. 
Fig.~\ref{fig:crossing}(f–h) show this distribution in the FM and PM phases, and at the MNP. 
In the FM phase, its center grows with system size, while in the PM phase it becomes sharply peaked at zero, consistent with earlier discussion.
At the MNP, all these measures cross, suggesting that the $\Delta F$ distribution becomes scale-invariant~\cite{Chalker2002NegativeScaling}.
Indeed, as seen in Fig.~\ref{fig:crossing}(h), the distribution at $p_c$ is scale invariant and displays a nonanalytic kink at $\Delta F=0$, arising from the one-to-one correspondence between configurations of opposite $\Delta F$.
Together with the Nishimori condition, this constrains the kink to satisfy $\partial_+ \mathbb P(0)+\partial_- \mathbb P(0) = \mathbb P(0)$, where $\mathbb P(x)$ is the probability density of $\Delta F$.

\begin{figure}[t]
    \centering
    \includegraphics[width=1.0\linewidth]{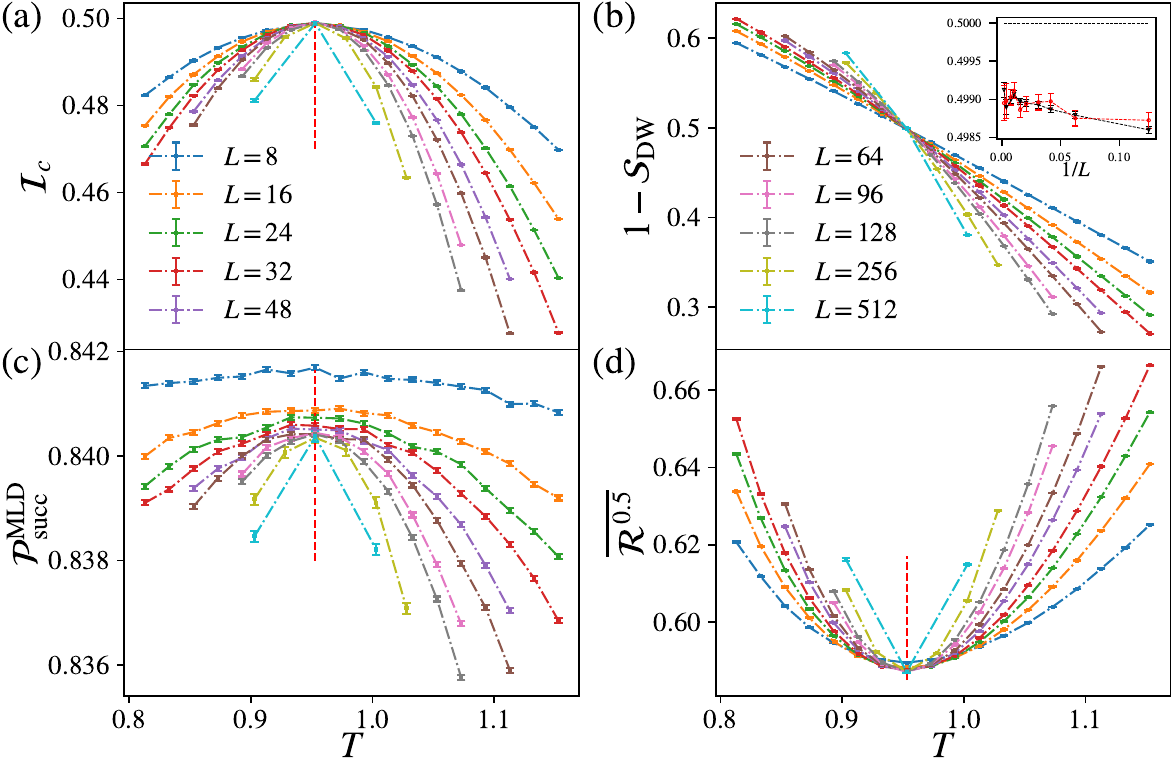}
    \caption{{\bf Behavior of quantities under perturbations away from the Nishimori line at $p=p_c$}.
    (a) $\mathcal I_c$; (b) $1$$-$$\mathcal{S}_\text{DW}$; (c) $\mathcal{P}_\text{succ}^\text{MLD}$; (d) $\overline{R^{0.5}}$.
    Quantities in (a), (c) and (d) reach an extremum at $T=T_c$, marked by the vertical dashed line.  
    The inset of panel (b) compares $\mathcal{I}_c$ (red) and $1$$-$$\mathcal{S}_\text{DW}$ (black) at $T_c$, with the latter exhibiting substantially smaller error bars.
    The crossing value, extrapolated to infinite size, deviates from $1/2$. 
    The legends are shared between panels.
    }
    \label{fig:tline}
 \end{figure}

\textit{Relevant perturbation away from the Nishimori line.---}
For general $\beta$, $\mathcal{I}_c,\mathcal{P}^\text{MLD}_\text{succ},\overline{\mathcal{R}^{0.5}}$ obey the inequalities in \Fig{fig:surface_mapping}(c).
Interpreting these as indicators of the phase transition implies that the phase boundary is vertical at the MNP, analogous to the FM–PM boundary argument in Ref.~\cite{Nishimori1981} but without any assumption. 
This necessitates that the second relevant direction at the MNP aligns with the temperature axis~\cite{Doussal1988locationMNP,Harris1989prbepsilonexpansion}.
As shown in Fig.~\ref{fig:tline}, all three quantities reach extrema at $T_c$, confirming the expected inequalities and illustrating the optimality of the MLD and Bayes decoders at Nishimori temperature~\cite{Nishimori93decoding, Nishimori2001book, Zdeborova16inferencethreshold}.
\Fig{fig:tline}(b) shows the domain-wall entropy $\mathcal {S}_\text{DW}$; 
the inset compares $1$$-$$\mathcal {S}_\text{DW}$ with $\mathcal{I}_c$ at the MNP, demonstrating reduced error for $1$$-$$\mathcal {S}_\text{DW}$, as they are related by the variance-reduction estimator described earlier. 
Although equivalent at the MNP, away from the Nishimori line, the domain-wall entropy (and other quantities, e.g., $\mathcal{P}^\text{Bayes}_\text{succ}$, $d_W$) do not exhibit extremal behavior, and crossings for different system sizes persist. 
This may seem at odds with the thermodynamic limit, where these quantities should peak at $T_c$, but it arises from slow scaling near the MNP.
Similar to magnetization, the sign of the spin correlation  $\overline{\text{sgn}\langle S_iS_j\rangle}$ is maximized on the Nishimori line, whereas the correlation itself is not for finite-size systems \cite{Nishimori2001book}.
The quantities with extremal properties clearly illustrate the reentrant nature~\cite{Nishimori25temperaturechaos} of the phase diagram, as shown in Fig.~\ref{fig:surface_mapping}(c).
Nevertheless, due to their extreme behavior, these quantities are insensitive to temperature variations and thus unsuitable for reliably extracting the thermal exponent $\nu_T$.
Moreover, the multicritical nature of the MNP and the small value of $1/\nu_T$ restrict the reliable analysis to a narrow window around $T_c = 1/\beta_{p_c}$ \cite{Chalker2002Mapping}. 
We therefore perform finite-size scaling using quantities such as $d_W, \mathcal{S}_\text{DW}$ within a narrow window, obtaining $1/\nu_T = 0.251(2)$, consistent with previous Monte Carlo results~\cite{Ettore2008MC, Ettore2009strong} (see App.~\ref{app:fss}).
$\mathcal S_\text{DW}$, like $\mathcal I_c$ on the Nishimori line, exhibits minimal finite-size effects under temperature perturbations.

\textit{Discussion and concluding remarks.---}
In this Letter, we revisit Nishimori multicriticality in the RBIM through the lens of quantum-information–inspired measures.
Using a fermionic transfer matrix approach, we obtain high-precision estimates of the critical point $p_c$ and associated critical exponents $\nu, \nu_T, \eta$. 
Remarkably, the coherent information and related quantities exhibit extremal behavior along the Nishimori line even for finite systems, reflecting the optimality of different decoding strategies and providing a natural certification for the decoder performance.
Viewed as statistical indicators, these quantities constrain the phase diagram without additional assumptions.
Our high-precision analysis shows that the critical points extracted from all these quantities coincide to within seven decimal places, which we attribute to the scale invariance of the domain wall free energy distribution, a natural assumption on the Nishimori line where no spin-glass phase exists.
These results highlight the advantages of quantum information based observables, which display significantly reduced finite-size effects compared with conventional indicators.
More general deformations of the model would render it beyond the {\it non-interacting} Majorana fermion solvable regime, such as the non-Clifford measurement or noise upon toric code that endows fermion interaction~\cite{GY2024prxq, Tarun2024Selfdual, Zhu19selfdual}.
The boundary-flip Monte Carlo~\cite{HasenbuschJPI1993,Hasenbusch1993PhysicaA,HukushimaPRE1999} or tensor-network techniques~\cite{Harrow22shallow, Zhang22sampling, GY2023prl, Ippoliti23unraveling, gy2025nishimoriuniversality, YoujinPRB2025} offer promising routes for evaluating these measures, provided truncation and nonlinear effects are carefully controlled.

Overall, this work demonstrates that information-theoretic observables offer a powerful and precise framework for probing multicriticality in disordered systems.
The behavior of these generalized information measures is expected to be universal for models satisfying the Nishimori relation.
The inequalities could potentially be generalized to the $Z_N$ random bond Potts model~\cite{Picco02randomPotts}, and to the higher dimensional RBIM or random plaquette gauge model~\cite{WANG200331}, the latter of which is relevant to the surface code with measurement errors~\cite{TQM}. 
The DWFE distribution is also applicable to error correction with post-selection \cite{chen2025postselection,lee2025efficientpostselectiongeneralquantum}.
We anticipate that these ideas will inspire further studies of quantum error correction, Nishimori physics, and more general phase transitions in open quantum systems with measurement or noise.

\textit{Acknowledgments.---}
We thank Hidetoshi Nishimori and Simon Trebst for valuable comments on our manuscript and Shang-Qiang Ning, Ze-Yao Han and Shiwei Zhang for helpful discussions.
GYZ thanks Andreas Ludwig, Simon Trebst, Romain Vasseur, Sam Garratt, Hidetoshi Nishimori for collaborations and discussions on the domain wall entropy and coherent information in criticality of open quantum systems.
GYZ acknowledges the support of National Natural Science Foundation of China-Young Scientists Fund (No. 12504181) and Start-up Fund of HKUST(GZ) (No. G0101000221) and Guangdong provincial project (No. 2024QN11X201). 
XDD acknowledges the support from the New Cornerstone Science Foundation. 
The Flatiron Institute is a division of the Simons Foundation.
The Pfaffian calculations utilize the \texttt{pfapack} library~\cite{wimmer2012algorithm}.

\bibliography{ref}

\newcommand{\Zbp}{{\mathcal{Z}_{\beta_p}}}       
\newcommand{\Zbpt}{{\mathcal{Z}'_{\beta_p}}} 
\newcommand{\Zb}{{\mathcal{Z}_{\beta}}} 
\newcommand{\Zbt}{{\mathcal{Z}'_{\beta}}} 
\newpage
\onecolumngrid        % leave two–column 
\newpage

\appendix
\section{End Matter}\label{app:end_matter}
\twocolumngrid
\textit{Properties along the Nishimori line---}
In the RBIM with disorder strength $p$ and random coupling $\{\tau_{ij}\}$, if an observable $O(\tau)$ is gauge-invariant, i.e.,
\eq{
O(\tau_{ij}) = O(\tau_{ij} \theta_i \theta_j), \ \ \forall \{\theta_i = \pm 1\}. 
}{}
Then, it follows from Refs.~\cite{Nishimori1981,Nishimori2001book} that the disorder average satisfies (we omit a constant prefactor for brevity)
\eq{\overline{O(\tau)} = \sum_{\tau} \mathcal{Z}(\tau, \beta_p) \, O(\tau),}{}
where $\mathcal {Z}(\tau, \beta_p)$ is the partition function at the inverse temperature $\beta_p = -\frac 1 2 \log\frac{p}{1-p}$. 
We refer to this identity as the Nishimori relation. 
For clarity, we denote the disorder configuration by $\tau=\{\tau_{ij}\}$, and  the corresponding configuration with twisted boundary conditions by $\tau'$. 
In the following, we will often omit the explicit dependence on $\tau$, using $\Zbp$ to denote $\mathcal{Z}(\tau,\beta_p)$ on the Nishimori line and $\Zb$ for $\mathcal{Z}(\tau,\beta)$ at a general inverse temperature $\beta$. When twisted boundary conditions are imposed, the corresponding partition function is denoted by $\Zbt$.

Using this relation, we establish several exact properties on the Nishimori line. We begin by proving the inequalities stated in the main text.

\begin{inequality}
The disorder-averaged quantity $\overline{\mathcal R^{0.5}(\beta)}\equiv\overline{\sqrt{\mathcal{Z}'(\tau,\beta)/\mathcal{Z}(\tau,\beta)}}$ attains its minimum on the Nishimori line:
\eq{
\overline{\mathcal R^{0.5}(\beta)}\geq \overline{\mathcal R^{0.5}(\beta_p)}.
}{}
\end{inequality}
\textit{Proof:}
\begin{equation}
\begin{split}
\overline{\mathcal R^{0.5}(\beta)}
&= \sum_\tau \Zbp \sqrt{\frac{\Zbt}{\Zb}} = \frac{1}{2} \sum_\tau \Zbp \sqrt{\frac{\Zbt}{\Zb}} + \Zbpt \sqrt{\frac{\Zb}{\Zbt}} \\
&= \frac{1}{2} \sum_\tau \sqrt{\Zbp \Zbpt} \left( \sqrt{\frac{\Zbt \Zbp}{\Zb \Zbpt}} + \sqrt{\frac{\Zb \Zbpt}{\Zbt \Zbp}} \right) \\
&\geq \sum_\tau \sqrt{\Zbp \Zbpt}=\sum_{\tau}\Zbp\sqrt{\frac{\Zbpt}{\Zbp}}
= \overline{\mathcal R^{0.5}(\beta_p)}.
\end{split}
\end{equation}
The inequality follows from the arithmetic–geometric mean inequality.

We can also establish the following inequalities, which are directly related to $\mathcal{I}_c, \mathcal{P}_\text{succ}^\text{MLD}$ introduced in the main text.
\begin{inequality}
Let $P(\tau)$ be a positive, gauge-invariant function of $\tau$ and $\tau'$ is the bond configuration after boundary condition flip (under twisted boundary condition). Then the following inequality holds:
\eq{\overline{\log\left(\frac{P(\tau)}{P(\tau)+P(\tau')}\right)}\leq  \overline{\log\left(\frac{\Zbp}{\Zbp+\Zbpt}\right)}.
}{}
\end{inequality}
\textit{proof:}
The inequality is equivalent to:
\eq{\overline{\log\left(\frac{P(\tau)}{\Zbp}\right)}\leq \overline{\log\left(\frac{P(\tau)+P(\tau')}{\Zbp+\Zbpt}\right)}.
}{}
Using the gauge invariance of $P(\tau)$, the LHS can be written as
\eq{
\begin{split}
&\text{LHS} = \sum_\tau \Zbp \log \left(\frac{P(\tau)}{\Zbp}\right)\\
&=\frac 1 2\sum_\tau \Zbp \log \left(\frac{P(\tau)}{\Zbp}\right) + \Zbpt \log \left(\frac{P(\tau')}{\Zbpt}\right) \\
&\leq \sum_\tau \frac{\Zbp+\Zbpt}{2} \log\left(\frac{P(\tau)+P(\tau')}{\Zbp+\Zbpt}\right)=\text{RHS},
\end{split}
}{}
where the inequality follows from Jensen's inequality. 
Thus, with $P(\tau) = \Zb$, we have $\mathcal{I}_c(\beta)$ is maximized on the Nishimori line
$\mathcal I_c(\beta)\leq \mathcal I_c(\beta_p)$.
\begin{inequality}
Let $P(\tau)$ be a positive, gauge-invariant function of $\tau$, $\tau'$ is the bond configuration after boundary condition flip.  
Then the following inequality holds:
\eq{\overline{\frac{P(\tau)}{P(\tau)+P(\tau')}}\leq  \overline{\Theta({\Zbp-\Zbpt})}.
}{seq:theorem3}
\end{inequality}
\textit{proof:} 
Using the gauge invariance of $P(\tau)$, the LHS can be written as
\eq{
\begin{split}
&\text{LHS} = \frac{1}{2}\sum_\tau  \Zbp \frac{P(\tau)}{P(\tau)+P(\tau')}+ \Zbpt\frac{P(\tau')}{P(\tau)+P(\tau')}\\
&\leq \frac{1}{2}\sum_\tau  \mathbf{max}(\Zbp,\Zbpt)=\sum_\tau \Zbp \Theta({\Zbp-\Zbpt})\\
&=\overline{\Theta({\Zbp-\Zbpt})}.
\end{split}
}{}

These inequalities admit a natural interpretation in quantum error correction: $P(\tau)$ can be viewed as the inference probability of a decoder, and gauge invariance must hold since the decoder depends only on the syndrome rather than the specific error configuration $\tau$. 
Inequality 2 implies that, when $\overline{\log \mathcal{P}_\text{succ}}$ is used as the score function, the Bayes decoder with $\beta = \beta_p$ is optimal. Inequality 3 further shows that, when $\overline{\mathcal{P}_\text{succ}}$ is used as the score function, the MLD with $\beta = \beta_p$ is optimal.\\
\textbf{Variance reduction using Nishimori relation.}
Next, we demonstrate how the Nishimori relation can be exploited to reduce statistical errors of observables computed along the Nishimori line.
For any gauge-invariant observable $O(\tau)$, one finds
\eq{
\begin{split}
   & \overline{O(\tau)} =\sum_{\tau}\frac{1}{2}\Zbp O(\tau)+\frac{1}{2}\Zbpt O(\tau')\\
    &=\sum_{\tau} \frac {\Zbp + \Zbpt}{2} \left( \frac{\Zbp}{\Zbp+\Zbpt} O(\tau) + \frac{\Zbpt}{\Zbp+\Zbpt}O(\tau')\right)\\
    &=\sum_{\tau} \Zbp \left( \frac{\Zbp}{\Zbp+\Zbpt} O(\tau) + \frac{\Zbpt}{\Zbp+\Zbpt}O(\tau')\right)\\
    &=\overline{\frac{\Zbp}{\Zbp+\Zbpt} O(\tau) + \frac{\Zbpt}{\Zbp+\Zbpt}O(\tau')}.
    \end{split}
}{eq:better_estimator}
Hence, the last expression provides an alternative unbiased estimator for $O(\tau)$ with significantly reduced variance.
For instance, taking $O$ as the log-posterior $\log_2{\frac{\Zbp}{\Zbp+\Zbpt}}$ defined in \Eq{eq:coherent_info}, we have 
\eq{
\overline{\log_2{\frac{\Zbp}{\Zbp+\Zbpt}}}=\overline{\frac{\Zbp}{\Zbp+\Zbpt} \log_2{\frac{\Zbp}{\Zbp+\Zbpt} } + \mathcal{Z}\leftrightarrow\mathcal{Z}'}.
}{eq:new_estimator}
Comparing Eqs.~(\ref{eq:coherent_info}) and (\ref{eq:dwentropy}) immediately gives the relation along the Nishimori line:
$\mathcal{I}_c(\beta_p)+\mathcal{S}_{\rm{DW}}(\beta_p)=1$.

While both estimators converge to the same expectation value in the infinite-sample limit, their statistical errors can differ substantially. 
This is evident for the estimation of $\mathcal{P}_\text{succ}^\text{MLD}$. 
The original estimator is a Heaviside step function taking only values 0 or 1. Using the improved estimator,
\eq{
\mathcal{P}_\text{succ}^\text{MLD}= \overline{ \frac{\mathbf{max}(\Zbp,\Zbpt)}{\Zbp+\Zbpt}}.
}{}
Now, the disorder-averaged value can take any value between 0 and 1, substantially reducing variance. This estimator effectively counts contributions from both $\tau$ and $\tau'$ configurations, increasing the effective sample size. Near the MNP, the variance of the coherent information is reduced by a factor of $\sim 2.6$. While one could further reduce variance by linearly combining the two estimators, in practice the improvement over \Eq{eq:new_estimator} is negligible. 
Consequently, we adopt \Eq{eq:new_estimator} for all calculations along the Nishimori line.

\textbf{\flushleft Kink condition of $\Delta F $ distribution.}
Next, we prove the kink relation for the distribution of $\Delta F\equiv \log(\Zbp/\Zbpt)$ at 0 (along the Nishimori line).
Define the probability density
$\mathbb{P}(x)\equiv\overline{\delta (\log(\Zbp/\Zbpt)-x)}$.
Since the delta function is gauge invariant, we can write
\eq{
\mathbb{P}(x)=\sum_{\tau} \Zbp\delta (\log(\frac{\Zbp}{\Zbpt})-x).
}{}
Because configurations with $\Delta F>0$ and $\Delta F <0$ are in one-to-one correspondence, we have
\eq{
\begin{split}
\mathbb{P}(x)&=\sum_{\tau} \Zbpt \delta (\log(\frac{\Zbpt}{\Zbp})-x)= \sum_{\tau} \Zbpt \delta (\log(\frac{\Zbp}{\Zbpt})+x)\\
&=e^{x} \sum_{\tau} \Zbp \delta (\log(\frac{\Zbp}{\Zbpt})+x)
= e^{x} \mathbb{P}(-x),
\end{split}
}{}
where the last equality follows from the standard property of the delta function.
From this symmetry, the left and right derivatives at the origin satisfy
$\partial_+\mathbb P(0) +\partial_-\mathbb P(0) = \mathbb P(0)$, which is the desired kink condition.

\textit{Fitting results of different quantities along the Nishimori line.---}
Around the critical point, a dimensionless observable $Q$ is expected to follow the finite-size scaling form of $Q  = f((p-p_c)L^{1/\nu})  + L^{-\omega} g((p-p_c)L^{1/\nu})$,
where $f(x),g(x)$ are analytic functions parameterized as polynomials $f(x) = \sum_{i=0}^{n_f-1} a_i x^i$ and $g(x) = \sum_{i=0}^{n_g-1} b_i x^i$.
Here, $\nu$ is the correlation-length critical exponent, and the term $L^{-\omega}$ represents the leading correction to scaling.
Table~\ref{tab:end_matter_res_tab} lists fits for various quantities along the Nishimori line, producing the results shown in Fig.~\ref{fig:crossing}; further details are given in App.~\ref{app:fss}. 
\onecolumngrid
\begin{table*}[]
    \centering
   \begin{tabular}{c|c|c|l|l|l|l|l|l}\hline
& $\chi^2/\text{dof}$ & $p_c$& $1/\nu$&$\omega$ &$a_0$& $a_1$& $a_2$&$b_0$ \\\hline
 $\mathcal I_c$
&104.37/102 &0.10922145(60) & 0.6510(13) & 2.1(19) & 0.498999(54) & -2.689(16) & 0.1(10) & -0.09(55)\\\hline
 $P_\text{succ}^{\text{MLD}}$
 &110.28/102 &0.10922114(95) & 0.6518(16) & 1.34(30) & 0.840392(43) & -1.0476(73) & -1.44(45) & 0.021(19)\\\hline
 $P_\text{succ}^{\text{Bayes}}$
&106.42/102 &0.1092215(10) & 0.6511(14) & 1.07(78) & 0.778509(69) & -1.3224(84) & -0.74(52) & 0.0040(86)\\\hline
 $d_W$
&98.67/102 &0.10922133(70) & 0.6542(12) & 1.369(82) & 2.24113(53) & -17.756(93) & 51.0(58) & -1.10(27)\\\hline
 $\overline{\mathcal R^{0.1}}$
&100.00/102 &0.10922140(71) & 0.6530(12) & 1.39(11) & 0.820264(39) & 1.2999(70) & -2.77(44) & 0.063(22)\\\hline
 $\overline{\mathcal R^{0.3}}$
&101.94/102 &0.10922148(72) & 0.6517(13) & 1.44(22) & 0.636971(69) & 2.320(13) & -2.67(81) & 0.074(49)\\\hline
  $\overline{\mathcal R^{0.5}}$
&102.57/102 &0.10922149(72) & 0.6514(13) & 1.48(30) & 0.587177(73) & 2.530(14) & -2.10(90) & 0.068(62)\\\hline
combined &--- &0.10922117(41) & 0.6523(12) &1.353(45)&---&---&---&--- \\\hline
 \end{tabular}
    \caption{Fitting results using different quantities along the Nishimori line with $n_f=3$ and $n_g=1$. 
    The fits are performed using data for system sizes ranging from $L_\text{min}=32$ to $L_\text{max}=512$.
    The last row gives the bootstrap fitting result of combining all these quantities with shared $p_c, 1/\nu$ and $\omega$. 
    See App.~\ref{app:fss} for more details of the fitting and for other choice of $L_\text{min}$.
    }
    \label{tab:end_matter_res_tab}
\end{table*}

\newpage

\onecolumngrid        % leave two–column 

\newpage

\section{Fermionic transfer matrix and the algorithm details.}\label{app:alg}

In this section, we give the details of the mapping from the random bond Ising model to the fermionic transfer matrix.
Using the transfer matrix representation in the spin basis, the partition function of this model can be written as 
\eq{
\mathcal Z(J,\beta)= \mathcal A\times   \bra{\psi_l}\prod_{n=0}^{L_x-1}\hat H_n\hat V_n\ket{\psi_r},
}{eq:partition_function}
where $\ket{\psi_l},\ket{\psi_r}$ depend on the boundary condition and will be discussed later. The transfer matrices are defined as:
\begin{equation}
\begin{split}
\hat H_n &= \exp(\sum_{i=0}^{L_y-1} \kappa(n,i)\sigma^x_i),\\
\hat V_n &= \exp(\sum_{i=0}^{L_y-1} \beta J_v(n,i) \sigma^z_i \sigma^z_{i+1})
\end{split}
\end{equation}
with $J_v,J_h = \pm 1$ representing the vertical (horizontal) bond coupling and  
\begin{equation}
\kappa(n,i) =
\begin{cases}
\kappa_0, & \text{if } J_h(n,i) = 1 \\
\kappa_0 + \mathrm{i} \dfrac{\pi}{2}, & \text{if } J_h(n,i) = -1
\end{cases},\kappa_0 = \text{arctanh}(e^{-2\beta}).
\end{equation}
With $N_\text{hbond}$ denoting the number of horizontal bonds, the normalization factor in \Eq{eq:partition_function} is given by
\begin{equation}
\mathcal A = \left(\frac{\sinh(2\kappa_0)}{2}\right)^{-\frac{N_\text{hbond}}{2}}\times (-\mi)^{N_{J_h=-1}}.
\end{equation}

Note that the transfer matrix effectively corresponds to a one-dimensional spin Hamiltonian, which can be mapped to fermions via the Jordan–Wigner transformation.
Here, we directly introduce $2L_y$ Majorana fermions $\gamma_{2i},\gamma_{2i+1}$ and define the spin operators as
$\sigma^z_{i} = (-\mathrm i)^i\gamma_0\gamma_1\gamma_2\cdots \gamma_{2i}$,
$\sigma^x_i = -\mathrm{i}\gamma_{2i}\gamma_{2i+1}$.
It is straightforward to verify that these definitions satisfy the Pauli algebra:
\begin{equation}
\begin{split}
[\sigma_i^x, \sigma_j^x] &= 0, \quad [\sigma_i^z, \sigma_j^z] = 0, \\
(\sigma_i^x)^2 &= 1, \quad (\sigma_i^z)^2 = 1, \\
\sigma_i^x \sigma_i^z &= -\sigma_i^z \sigma_i^x, \\
[\sigma_i^x, \sigma_j^z] &= 0 \quad (i \ne j).
\end{split}
\end{equation}
In the Majorana fermion representation, the interaction term in the transfer matrix $\hat V_n$ takes the form:
\begin{equation}
\sigma_i^z \sigma_{i+1}^z = 
\begin{cases}
-\mathrm{i} \, \gamma_{2i+1} \gamma_{2i+2}, & \text{for } i < L_y-1, \\
\mathrm{i} \, \mathcal{P} \, \gamma_{2L_y-1} \gamma_0, & \text{for } i = L_y-1.
\end{cases}
\end{equation}
Here, $\mathcal{P} = \prod_i \sigma_i^x$ is the parity operator (or the $\mathbb{Z}_2$ charge) of the model, reflecting the model’s underlying $\mathbb{Z}_2$ symmetry.
Inserting these expressions back into \Eq{eq:partition_function}, we obtain
\eq{
\begin{split}
\hat H_n &= \exp(-\mathrm i \sum_{i=0}^{L_y-1} \kappa(n,i)\gamma_{2i}\gamma_{2i+1}),\\
\hat V_n &= \exp(-\mathrm i \sum_{i=0}^{L_y-2} \beta J_v(n,i) \gamma_{2i+1} \gamma_{2i+2} -\mathrm i  \beta J_v(n,L_y-1) \mathcal P\gamma_{0} \gamma_{2L_y-1}).
\end{split}
}{}

Now, we turn to the boundary wavefunctions, which depend on the imposed boundary conditions. For the free boundary condition, all spin configurations are allowed. Consequently, the boundary state can be written as a superposition of all states: 
\begin{equation}
    \psi_\text{free} = \sum_{\vect s} \ket{s_0s_1s_2\cdots s_{L_y-1}}
\end{equation}
This state is an eigenstate of all operators ${\sigma_i^x}=1$.
It has even parity $\mathcal P = 1$ and thus corresponds to the ground state of the Hamiltonian $H=\sum \sigma^x_i=\sum \mi\gamma_{2i}\gamma_{2i+1}$.
The other physical boundary condition is the fixed boundary condition with spin all up or all down.
For this boundary condition, we can use a GHZ state:
\begin{equation}
\psi_\text{fixed} = \ket{0000\cdots0}  + \ket{1111\cdots1}.
\end{equation}
This state has parity $\mathcal P =1$ and is stabilized by the set of operators $\{Z_{i}Z_{i+1}(i<L_y-1), \mathcal P \}$.
Consequently, it is the ground state of the Hamiltonian $H = - \mi \gamma_{2L_y-1}\gamma_0 + \mi \sum_{i=0}^{L_y-2}\gamma_{2i+1}\gamma_{2i+2} $.

Now, we turn to the problem of calculating the overlap in \Eq{eq:partition_function}. 
As long as both $\ket{\psi_l}$ and $\ket{\psi_r}$ have non-negative amplitudes in the computational basis (i.e., the $Z$-basis), the overlap can be interpreted as a classical partition function with corresponding boundary conditions. 
In this case, the overlap is manifestly \emph{positive definite}, reflecting the absence of a sign problem. 
In the Majorana representation, the overlap is generally expressed as a Pfaffian. 
Nevertheless, given that we know a priori that the overlap has no sign ambiguity, the calculation can be dramatically simplified. 
This simplification is inherited from Majorana quantum Monte Carlo simulations. 
To compute the overlap, we introduce a replica of the system with Majorana operators 
$\gamma_{i}^1$ and $\gamma_i^2$ for $i = 0, \cdots, L_y{-}1$. 
Then the overlap in the replicated (double) Hilbert space becomes
\begin{equation}
\mathcal Z(J,\beta)^2= \mathcal A^2 \times   (\bra{\psi^1_l}\otimes \bra{\psi^2_l})\prod_n\hat H^1_n\hat H^2_n\hat V^1_n\hat V^2_n(\ket{\psi^1_r}\otimes \ket{\psi^2_r}).
\end{equation}
Now, we define the combined transfer matrix operators in the replicated space:
\begin{equation}
\begin{split}
\hat {\mathcal H}_n \equiv \hat {H}^1_n\hat {H}^2_n&= \exp\left(-\mathrm \sum_{i=0}^{L_y-1} \kappa(n,i) (\mi\gamma^1_{2i}\gamma^1_{2i+1}+\mi\gamma^2_{2i}\gamma^2_{2i+1}) \right),\\
\hat {\mathcal V}_n\equiv \hat {V}^1_n\hat {V}^2_n &= \exp\left(-\mathrm i \sum_{i=0}^{L_y-2} \beta J_v(n,i) \gamma^1_{2i+1} \gamma^1_{2i+2} -\mathrm i  \beta J_v(n,L_y-1) \mathcal P\gamma^1_{0} \gamma^1_{2L_y-1} + 1\rightarrow 2\right),
\end{split}
\end{equation}
where $(1 \rightarrow 2)$ denotes the same terms for the second replica with a superscript $2$.
By introducing complex fermion operators $c_i =\frac 1 2( \gamma_i^1 + \mi \gamma_i^2), c_i^\dagger = \frac 1 2 (\gamma_i^1 -\mi \gamma_i^2)$,
the Majorana bilinears combine as $\mi \gamma_i^1\gamma_j^1+ \mi \gamma_i^2\gamma_j^2= 2\mi(c^\dagger_i c_j - c_j^\dagger c_i)$.
Then we have
\begin{equation}
   \begin{split}
\hat {\mathcal H}_n&= \exp\left(-2\mi\mathrm \sum_{i=0}^{L_y-1} \kappa(n,i) c^\dagger_{2i}c_{2i+1} + \text{h.c.}\right),\\
\hat {\mathcal V}_n&= \exp\left(-2\mathrm i \sum_{i=0}^{L_y-2} \beta J_v(n,i) c^\dagger_{2i+1} c_{2i+2}  -2\mathrm i  \beta J_v(n,L_y-1) \mathcal Pc^\dagger_{0} c_{2L_y-1}  + \text{h.c.}\right).
\end{split}
\end{equation}
In the following, we assume that both $\ket{\psi_l}$ and $\ket{\psi_r}$ have even fermion parity so that the parity operator $\mathcal{P}$ can be replaced by 1. For the left and right wavefunctions in the replica space, note that if $\ket{\psi^1}$ is the ground state of a free-fermion Hamiltonian $\hat H^1 = \mathrm{i} \sum h_{ij} \gamma^1_i \gamma^1_j$, then the product state $\ket{\psi^1} \otimes \ket{\psi^2}$ is the ground state of a complex fermion Hamiltonian $\hat H = \mi \sum h_{ij} c_i^\dagger c_j + \text{h.c.}$.
The entire problem thus reduces to the well-known framework of propagating Slater determinants using fermionic Gaussian operators with a conserved $U(1)$ charge symmetry.
It can be shown that the overlap can be expressed as
\begin{equation}\label{eq:det_overlap}
    \bra{\psi_l}\prod_n \hat H_n \hat V_n \ket{\psi_r}
    = \pm \sqrt{ \left| \det \left( P_l^\dagger \prod_n \mathbb{H}_n \mathbb{V}_n P_r \right) \right| },
\end{equation}
where $\mathbb{H}_n$ and $\mathbb{V}_n$ are the matrix representations of the fermionic Gaussian operators $\hat H_n$ and $\hat V_n$, respectively. The matrices $P_l$ and $P_r$ represent the Slater determinant wavefunctions of the left and right boundary states $\ket{\psi_l}$ and $\ket{\psi_r}$.
The expression in \Eq{eq:det_overlap} is widely used in zero-temperature determinant quantum Monte Carlo simulations.
A corresponding formula also exists for periodic boundary conditions, where the overlap is replaced by a trace, resembling the finite-temperature algorithm. 
In this case, however, the parity operator $\mathcal P$ cannot be replaced by a constant, and explicit parity projection is required, which substantially increases the computational complexity.
Moreover, finite-temperature formulations are known to suffer from stronger numerical instabilities and higher computational cost.
For these reasons, we employ free or fixed boundary conditions to remain in the zero-temperature regime, where efficient and numerically stable determinant evaluation algorithms are well established in the literature.
To further reduce the computational complexity, we perform a basis transformation defined by $c_{2i} = \tilde{c}_{2i}$ and $c_{2i+1} = \mi \tilde{c}_{2i+1}$, which eliminates the need for complex arithmetic.

Overall, the algorithm of calculating the partition function is as follows:

\begin{algorithm}[H]
\caption{Computation of the Partition Function}
\KwIn{$P_r$, $P_l$, matrices $\mathbb{H}_n$, $\mathbb{V}_n$ in $\tilde c$ basis.}
\KwOut{Logarithm of the partition function}

$Q_r \leftarrow P_r$, $Q_l \leftarrow P_l$\;
$c_r \leftarrow 0$, $c_l \leftarrow 0$\;

\For{$i \leftarrow 0$ \KwTo $L_x/2 - 1$}{
    $Q_l \leftarrow \mathbb{V}_{L_x - i - 1} \mathbb{H}_{L_x - i - 1} Q_l$\;
    $Q_r \leftarrow \mathbb{H}_i \mathbb{V}_i Q_r$\;
    
    \If{$(i+1) \bmod n_\mathrm{stab} = 0$}{
        $(Q_l, R_l) \leftarrow \text{QR}(Q_l)$\;
        $c_l \leftarrow c_l +\sum_j \log\left(  |\text{Diag}(R_l)_j| \right)$\;
        
        $(Q_r, R_r) \leftarrow \text{QR}(Q_r)$\;
        $c_r \leftarrow c_r + \sum_j\log\left( |\text{Diag}(R_r)_j| \right)$\;
    }
}
\Return $c_l + c_r + \log\left[ |\det(Q_l^\dagger Q_r) |\right]$ + $\log|\mathcal A|$\;
\end{algorithm}
Here, $n_\text{stab}$ denotes the stabilization interval, which is chosen to ensure the numerical stability of the algorithm. A detailed analysis of this parameter will be provided later.
The algorithm involves both matrix multiplications and QR decompositions. Each step scales linearly with $L_x$, the system length in the propagation direction. Specifically, the matrix multiplications cost $O(L_y^2)$ per step because the matrices $\mathbb{H}_i$ and $\mathbb{V}_i$ are block-diagonal with block size proportional to $L_y$. In contrast, the QR decompositions, which are crucial for maintaining numerical stability, scale as $O(L_y^3)$ and thus dominate the overall computation. Consequently, the total time complexity of the algorithm is $O(L_x L_y^3)$.
By selecting a moderate value for $n_\text{stab}$ and leveraging modern computational resources, the algorithm remains highly efficient and can be scaled to large system sizes. This enables the extensive numerical investigations presented in this work.

\begin{figure}
    \centering
    \includegraphics[width=0.45\linewidth]{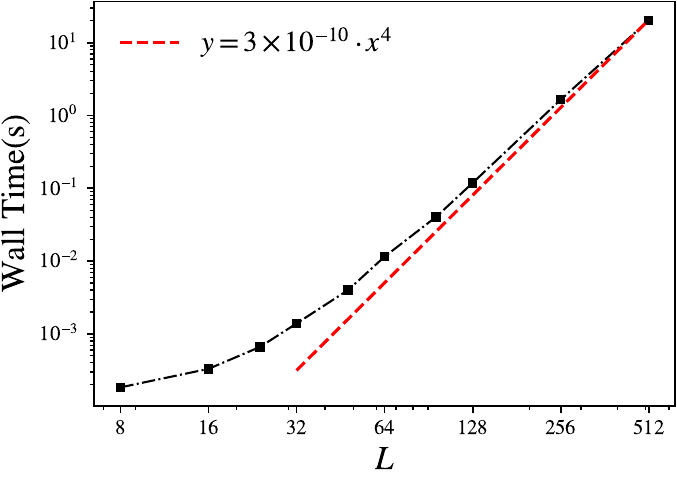}
    \caption{Computational cost of the algorithm. The figure shows the wall-clock time required to compute a single disorder realization—specifically, the outputs $\log Z$ and $\log Z'$—using one core of an AMD EPYC 7742 CPU.
    The red line corresponds to $y=3\times 10^{-10} x^4$, illustrating the asymptotic scaling behavior of the algorithm.
    }
    \label{fig:enter-label}
\end{figure}

In addition to the partition function, physical observables can also be computed using the fermionic transfer matrix formalism. In the following, we demonstrate how to evaluate the spin–spin correlation function at a given slice.

The spin-spin correlation can be expressed in terms of Majorana fermions as
\begin{equation}
    \braket{\sigma^z_i \sigma^z_j} =  \braket{(-\mi)^{j-i} \gamma_{2i+1}\gamma_{2i+2}\cdots \gamma_{2j}}
\end{equation}
To simplify the expression and eliminate the imaginary unit, we introduce a modified Majorana basis: $\gamma'_{2i} = \gamma_{2i}, \gamma'_{2i+1}=-\mi \gamma_{2i+1}$.
We then define the Green’s function matrix in the $\gamma'$ basis as $\gamma'$, $\mathcal G_{ij}=\braket{\gamma'_i\gamma'_j}$, which remains a real skew-symmetric matrix.
Using Wick’s theorem, the spin–spin correlation function reduces to the Pfaffian of a submatrix of $\mathcal G$:
\begin{equation}
   \braket{\sigma^z_i \sigma^z_j} = \braket{\gamma'_{2i+1}\gamma'_{2i+2}\cdots \gamma'_{2j+2}} = \text{pf}(\mathcal G_{2i+1:2j,2i+1:2j})
\end{equation}
Using the transformations from $\gamma$ to $\gamma'$ and from $c$ to $\tilde{c}$, the correlation matrix $\mathcal{G}$ can be calculated as
\begin{equation}
    \mathcal G_{ij} = \braket{\gamma'_i\gamma'_j}= 2(-1)^i\braket{\tilde c_i^\dagger \tilde c_j} - \mathbb I.
\end{equation}

The calculation of spin–spin correlations involves evaluating the Pfaffian of a skew-symmetric matrix, which has a computational complexity of $O(N^3)$. Consequently, computing the correlation between a single pair of spins separated by a distance $R$ requires $O(R^3)$ operations. In our study, we restrict the evaluation to correlations at the central slice ($n_x = L_x/2$), specifically between the boundary and bulk spins. As a result, the total cost scales as $O(L_y^4)$, which remains comparable to the overall algorithmic complexity.
We compute the Pfaffians using the \texttt{pfapack} library \cite{wimmer2012algorithm}, which implements the Parlett–Reid algorithm. In practice, the spin–spin correlation calculation accounts for less than 10\% of the total computational time.

It is worth noting that the spin–spin correlation always corresponds to a string (chain) operator, and its structure may allow for further optimization. 
In principle, information from previously computed Pfaffians could be reused to reduce computational overhead. For instance, a version of the Parlett–Reid algorithm without pivoting could be more efficient. 
However, pivoting remains essential in general to guarantee numerical stability in Pfaffian evaluation. 
We leave a systematic exploration of these potential optimizations to future work.

\section{Rounding error analysis.}

\begin{figure}
    \centering
    \includegraphics[width=0.7\linewidth]{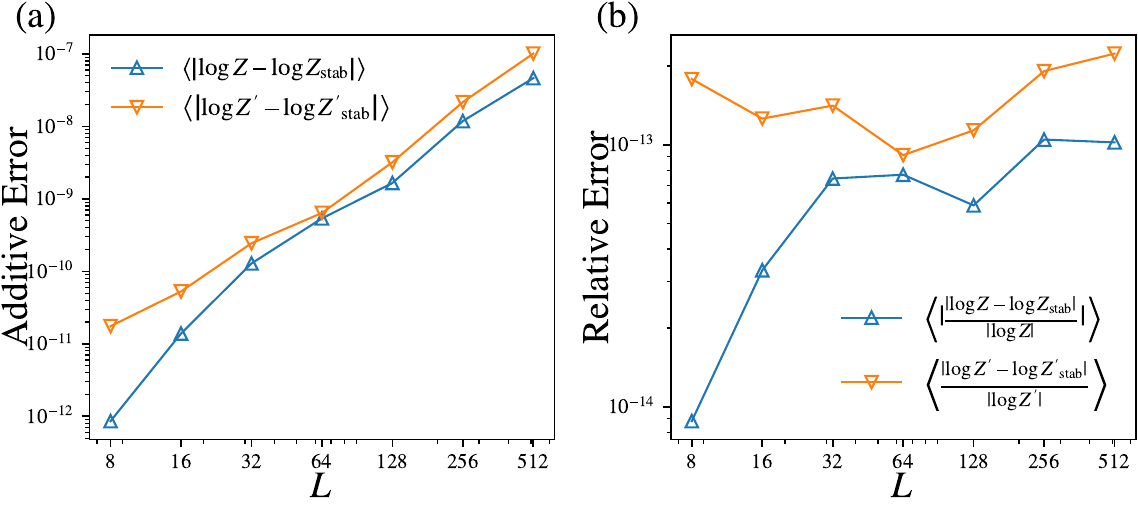}
    \caption{Error analysis at Nishimori point for different system sizes $L_x=L_y=L$ with fixed stabilization step $n_\text{stab} =5$. 
    Panels (a),(b) represent the additive error and relative error of $\log Z$ respectively. 
    The reference value $\log Z_\text{stab}$ is calculated by using $n_\text{stab}=1$.
    The data here is the average of 1000 disorder samples.
    }
    \label{fig:error}
\end{figure}
The numerical instability arises from repeated matrix multiplications in the algorithm. 
Specifically, the product $\prod_i \mathbb H_i\mathbb V_i$ tends to develop eigenvalues that grow or decay exponentially with the number of multiplications. The characteristic rate of this exponential behavior is known as the Lyapunov exponent, which has also been used to identify phase transitions in previous studies.
As the system size \( L_x \) increases, the transfer matrix accumulates a spectrum with increasingly extreme eigenvalues—ranging from exponentially large to exponentially small. 
This wide dynamic range inevitably exceeds the limits of floating-point precision, even in double precision, leading to significant rounding errors and numerical instability. This issue has long been recognized in both transfer matrix methods and determinant quantum Monte Carlo studies.
To address this, early approaches employed \textit{modified Gram-Schmidt} reorthogonalization to maintain numerical stability during matrix multiplication. Later developments in zero-temperature DQMC demonstrated that \textit{Householder QR decomposition} provides improved numerical stability.
In this work, we adopt the QR decomposition technique to stabilize the matrix multiplication process. 
Additionally, by using free or fixed boundary conditions, the number of vectors involved in the stabilization procedure is reduced from \( 2L_y \) (as required in standard Lyapunov exponent calculations) to \( L_y \), resulting in a roughly $4\times$ improvement in computational efficiency.
In the calculation shown in our paper, we choose $n_\text{stab} = 5$ for the Nishimori line and $n_\text{stab} = 4$ for the vertical temperature line. 
In \Fig{fig:error}, we demonstrate the rounding errors for different system sizes.
We observe that for system sizes up to $512 \times 512$, the average additive error in $\log \mathcal Z$ remains on the order of $\sim 10^{-7}$, while the relative error stays as low as $\sim 10^{-13}$.
These numerical errors are significantly smaller than the statistical uncertainties in the measured observables and therefore have a negligible impact on our results.

\section{Simulation and finite-size scaling details.}
\label{app:fss} 

\begin{figure*}
    \centering
    \includegraphics[width=1.0\linewidth]{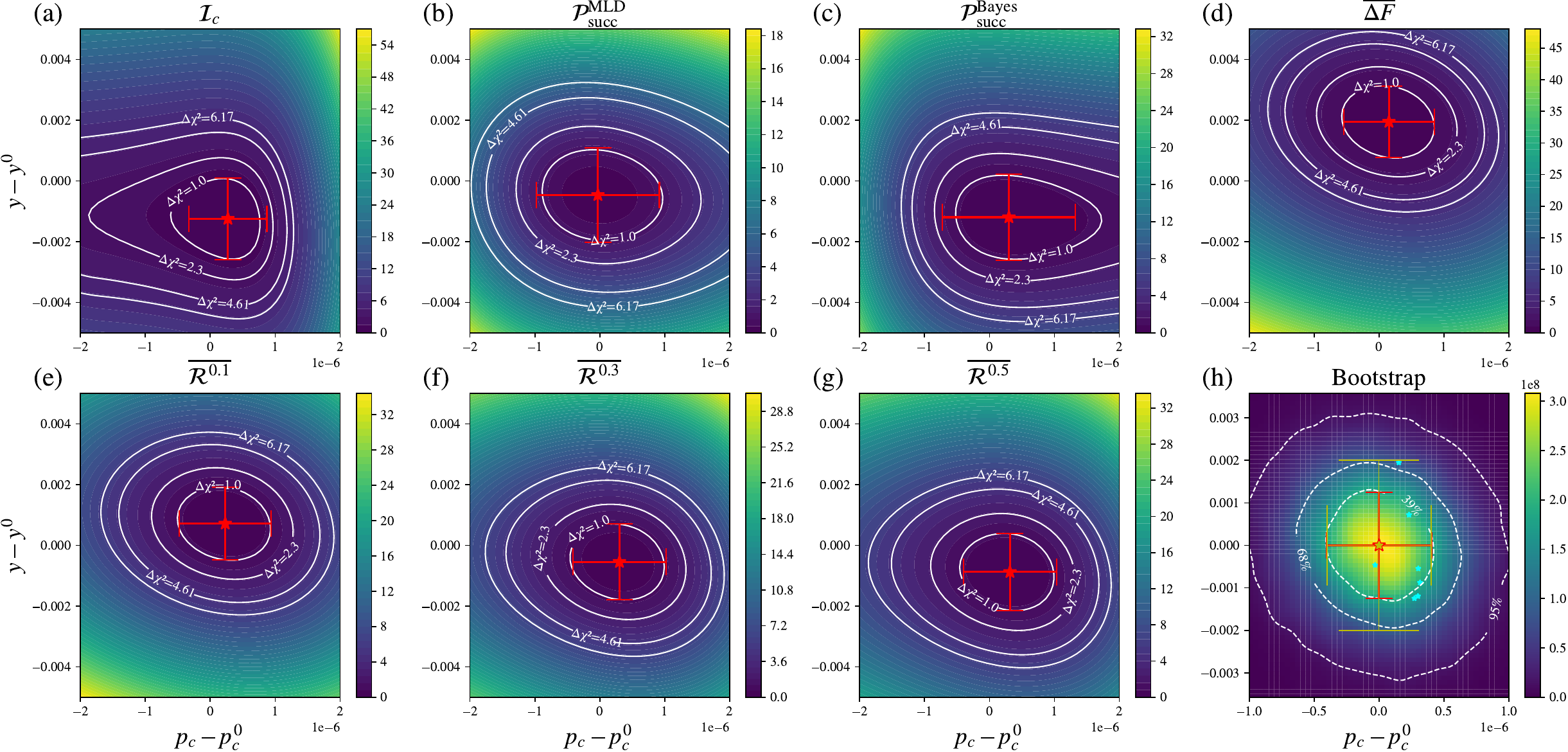}
    \caption{\textbf{Profile $\Delta\chi^2$ contour plot for individual fits and bootstrap results for the combined fit.} 
    (a--g) Heatmaps of $\Delta \chi^2$ for individual fits across various quantities. 
    Here, $\Delta \chi^2$ represents the difference between the \emph{re-optimized} $\chi^2$ values obtained by tuning $p_c$ and $y$. 
    The contours correspond to fixed $\Delta \chi^2$ levels indicated in each panel. 
    Red error bars denote the best-fit results obtained directly from least-squares fitting. (h) Bootstrap results for the combined fit. 
    The raw data are first divided into 10{,}000 bins, followed by bootstrap resampling of these bins. 
    All observables are computed using the same resampled bins to account for correlations between different quantities. 
    A total of 50{,}000 bootstrap samples are used. 
    The contours indicate regions enclosing different probability levels. 
    The red error bar marks the 68\% percentile confidence level, while the yellow point denotes the final reported value, incorporating all individual fit results (the points).
    The data shown correspond to Table~\ref{tab:nline_res} with $L_{\text{min}} = 32$. The $p_c^0, y^0$ used here are the fitting results from bootstrap.
    }
    \label{fig:S3}
\end{figure*}

\textit{Details of the simulation and finite-size scaling results along the Nishimori line.}
In Table~\ref{tab:nline_params}, we list the simulation parameters that are used along the Nishimori line, e.g. in \Fig{fig:crossing}.
\begin{table}[h]
 \setlength{\tabcolsep}{12pt}
    \centering
    \begin{tabular}
    {c|c|c|c|c}\hline
          $L$& $p_\text{min}$&$p_\text{max}$&  $\Delta p$ &$n_\text{sample}$\\\hline
          8-32& 0.108900&0.109550& 
    0.000050 &$6\times 10^7$\\\hline
 48,64& 0.108975& 0.109450&0.000025 &$6\times 10^7$\\\hline
 96,128& 0.109025& 0.109400& 0.000025&$6\times 10^7$\\\hline
 192& 0.109100& 0.109325& 0.000025&$6\times 10^7$\\\hline
 256& 0.109100& 0.109325& 0.000025&$1.2\times 10^7$\\\hline
 512& 0.109150& 0.109250& 0.000050&$1.2\times 10^7$\\ \hline
  \end{tabular}
    \caption{Simulation parameters for calculations on the Nishimori line.}
    \label{tab:nline_params}
\end{table}
Around the critical point, a dimensionless observable $Q$ is expected to follow the finite-size scaling form:
\eq{
Q  = f((p-p_c)L^{1/\nu})  + L^{-\omega} g((p-p_c)L^{1/\nu}),
}{eq:fss}
where $f(x),g(x)$ are analytic functions, which we parameterize as polynomials $f(x) = \sum_{i=0}^{n_f-1} a_i x^i$ and $g(x) = \sum_{i=0}^{n_g-1} b_i x^i$.
Here, $\nu$ is the correlation-length critical exponent, and the term $L^{-\omega}$ represents the leading correction to scaling.
Table~\ref{tab:nline_res} summarizes the fits to Eq.~\eqref{eq:fss} with $n_f = 3, n_g=1$ for various $L_\text{min}$ and observables along the Nishimori line. 
The goodness-of-fit values $\chi^2/\text{dof}$ are all close to 1, indicating that the scaling form is appropriate.
As $L_\text{min}$ increases from $8$ to $32$, the fitted parameters for different quantities converge.
The critical point $p_c$ agrees across the observables to high precision.
The exponent $\nu$ is also consistent among the observables, except that the value extracted from $d_W$ is slightly larger, which we attribute to stronger finite-size corrections in the domain-wall free energy.
In particular, the quantum-information–related observables such as $\mathcal I_c$, $\mathcal{P}_\text{succ}$ exhibit small finite-size corrections (i.e., the fitted $b_0$ values are small).
We also perform a simultaneous fit of all these quantities with shared parameters $p_c,1/\nu,\omega$ ($L_\text{min}=32$).
The combined fit incorporates information from all observables, leading to a more robust estimate of the critical parameters.
However, correlations exist among different quantities, and these must be properly taken into account.
To capture such correlations, we employ a bootstrap analysis on the raw simulation data.
For each $p,L$ datapoint, the original simulation data are first divided into 10,000 bins, and bootstrap resampling is then performed on these bins.
The observables are computed using the \emph{same} resampling bins, ensuring that correlations between different quantities are preserved.
For 50,000 bootstrap samples, we perform a combined fit in which $p_c, 1/\nu, \omega$ are shared parameters across all observables.
The 68\% percentile confidence intervals of the bootstrap distributions yield the results listed in the last row of Table~\ref{tab:nline_res}.
To further assess the robustness of our fitting procedure, Fig.~\ref{fig:S3} shows the \emph{re-optimized} $\Delta \chi^2$ contour plots for individual fits as well as the bootstrap distributions in the $p_c-y$ plane ($1/\nu$ is denoted as y).
The re-optimized $\Delta \chi^2$ contours are consistent with the least-squares fitting results, and the bootstrap distributions agree with the corresponding percentile error bars.
We adopt the estimates from the combined bootstrap fitting as our final results and enlarge the quoted uncertainties to cover all values obtained from individual fits.
\eq{
p_c = 0.1092212(4),\quad 1/\nu = 0.652(2).
}{eq:final_estimate}

\begin{table}[]
    \centering
    \begin{tabular}{c|c|c|c|c|c|c|c|c|c}\hline
&$L_\text{min}$&  $\chi^2/\text{dof}$ & $p_c$& $1/\nu$&$\omega$ &$a_0$& $a_1$& $a_2$&$b_0$ \\\hline
 $\mathcal I_c$
&8 &144.12/144 &0.10922157(34) & 0.64940(95) & 1.52(17) & 0.498994(25) & -2.709(11) & 0.01(98) & -0.0110(36)\\\hline
&16 &133.78/130 &0.10922134(58) & 0.6491(10) & 1.18(52) & 0.499019(59) & -2.713(12) & -0.1(10) & -0.0048(59)\\\hline
&24 &120.43/116 &0.10922119(91) & 0.6493(12) & 0.9(10) & 0.49904(13) & -2.710(14) & 0.0(10) & -0.0025(62)\\\hline
&32 &104.37/102 &0.10922145(60) & 0.6510(13) & 2.1(19) & 0.498999(54) & -2.689(16) & 0.1(10) & -0.09(55)\\\hline
 $P_\text{succ}^{\text{MLD}}$
&8 &156.85/144 &0.10922232(48) & 0.6508(11) & 1.171(27) & 0.840343(16) & -1.0526(51) & -1.47(44) & 0.01439(67)\\\hline
&16 &139.75/130 &0.10922133(60) & 0.6502(12) & 1.342(79) & 0.840387(22) & -1.0555(55) & -1.52(45) & 0.0219(42)\\\hline
&24 &126.30/116 &0.10922092(73) & 0.6500(13) & 1.48(18) & 0.840405(29) & -1.0564(63) & -1.49(45) & 0.032(16)\\\hline
&32 &110.28/102 &0.10922114(95) & 0.6518(16) & 1.34(30) & 0.840392(43) & -1.0476(73) & -1.44(45) & 0.021(19)\\\hline
 $P_\text{succ}^{\text{Bayes}}$
&8 &147.49/144 &0.10922172(43) & 0.6495(10) & 1.203(67) & 0.778502(18) & -1.3324(58) & -0.80(51) & 0.00696(82)\\\hline
&16 &136.24/130 &0.10922136(54) & 0.6492(11) & 1.39(21) & 0.778522(24) & -1.3343(64) & -0.84(52) & 0.0110(56)\\\hline
&24 &123.06/116 &0.10922112(63) & 0.6493(12) & 1.62(49) & 0.778535(29) & -1.3334(72) & -0.79(52) & 0.022(31)\\\hline
&32 &106.42/102 &0.1092215(10) & 0.6511(14) & 1.07(78) & 0.778509(69) & -1.3224(84) & -0.74(52) & 0.0040(86)\\\hline
 $d_W$
&8 &152.75/144 &0.10922038(35) & 0.65571(88) & 1.2673(78) & 2.24190(19) & -17.640(67) & 48.3(58) & -0.821(11)\\\hline
&16 &119.89/130 &0.10922089(45) & 0.65396(89) & 1.298(22) & 2.24153(29) & -17.779(68) & 49.8(55) & -0.885(47)\\\hline
&24 &106.99/116 &0.10922098(59) & 0.65362(99) & 1.306(47) & 2.24146(43) & -17.806(78) & 50.7(56) & -0.91(12)\\\hline
&32 &98.67/102 &0.10922133(70) & 0.6542(12) & 1.369(82) & 2.24113(53) & -17.756(93) & 51.0(58) & -1.10(27)\\\hline
 $\overline{\mathcal R^{0.1}}$
&8 &141.88/144 &0.10922081(34) & 0.65327(86) & 1.284(10) & 0.820227(14) & 1.2986(48) & -2.65(42) & 0.04619(86)\\\hline
&16 &123.13/130 &0.10922106(46) & 0.65212(91) & 1.304(30) & 0.820240(22) & 1.3053(52) & -2.70(42) & 0.0484(36)\\\hline
&24 &110.41/116 &0.10922106(61) & 0.6521(10) & 1.303(66) & 0.820240(32) & 1.3057(59) & -2.75(42) & 0.0483(87)\\\hline
&32 &100.00/102 &0.10922140(71) & 0.6530(12) & 1.39(11) & 0.820264(39) & 1.2999(70) & -2.77(44) & 0.063(22)\\\hline
 $\overline{\mathcal R^{0.3}}$
&8 &139.75/144 &0.10922126(35) & 0.65079(88) & 1.312(19) & 0.636942(24) & 2.3301(89) & -2.56(78) & 0.0485(17)\\\hline
&16 &128.53/130 &0.10922124(49) & 0.65022(96) & 1.306(59) & 0.636939(40) & 2.3360(98) & -2.55(80) & 0.0477(67)\\\hline
&24 &115.35/116 &0.10922114(66) & 0.6504(11) & 1.28(13) & 0.636929(62) & 2.334(11) & -2.62(80) & 0.045(15)\\\hline
&32 &101.94/102 &0.10922148(72) & 0.6517(13) & 1.44(22) & 0.636971(69) & 2.320(13) & -2.67(81) & 0.074(49)\\\hline
  $\overline{\mathcal R^{0.5}}$
&8 &140.66/144 &0.10922138(35) & 0.65014(90) & 1.328(26) & 0.587155(26) & 2.5451(99) & -1.99(87) & 0.0405(20)\\\hline
&16 &130.25/130 &0.10922128(50) & 0.64972(98) & 1.304(81) & 0.587145(45) & 2.550(11) & -1.96(89) & 0.0381(75)\\\hline
&24 &116.91/116 &0.10922116(68) & 0.6499(11) & 1.26(17) & 0.587131(71) & 2.547(12) & -2.04(90) & 0.034(16)\\\hline
&32 &102.57/102 &0.10922149(72) & 0.6514(13) & 1.48(30) & 0.587177(73) & 2.530(14) & -2.10(90) & 0.068(62)\\\hline

combined &32 &--- &0.10922117(41) & 0.6523(12) &1.353(45)&---&---&---&--- \\\hline
 \end{tabular}
    \caption{
    {\bf MNP perturbed along Nishimori line}.    
    Fitting results using different quantities along the Nishimori line with $n_f=3$ and $n_g=1$. 
    The fits are performed using data for system sizes ranging from $L_\text{min}$ to $L_\text{max}=512$.
    The last row gives the combined fitting results using all the above quantities with shared $p_c, 1/\nu, \omega$ and bootstrap method.
    }
    \label{tab:nline_res}
\end{table}

\begin{table}[]
    \centering
    \begin{tabular}{c|c|c|c|c|c|c}\hline
$L_\text{min}$&  $\Delta (\omega = 1)$ &$\chi^2/\text{dof}$ &$\Delta (\omega = 1.3)$ &$\chi^2/\text{dof}$ & $\Delta (\omega = 1.5)$ &$\chi^2/\text{dof}$  \\
          \hline
8 & 0.089754(25) & 259.68/145 & 0.089180(19) & 196.67/145 & 0.088930(21) & 277.09/145 \\\hline
16 & 0.089545(33) & 150.22/131 & 0.089219(27) & 147.36/131 & 0.089074(26) & 158.03/131 \\\hline
24 & 0.089448(46) & 127.37/117 & 0.089219(38) & 132.36/117 & 0.089116(35) & 137.85/117 \\\hline
32 & 0.089434(61) & 114.23/103 & 0.089244(51) & 118.15/103 & 0.089158(46) & 121.50/103 \\\hline
48 & 0.089463(94) & 91.40/89 & 0.089313(77) & 91.88/89 & 0.089245(69) & 92.37/89 \\\hline
64 & 0.08950(13) & 72.23/69 & 0.08937(11) & 72.89/69 & 0.089315(99) & 73.35/69 \\\hline
 \end{tabular}
    \caption{Fitting results for the central magnetization $m_\text{center}$ along the Nishimori line, using $n_f=3$ and $n_g=2$ ($n_g=1$ yields much larger $\chi^2/\text{dof}$). 
    $p_c,1/\nu$ in the scaling form \Eq{eq:m_scaling} are fixed to the values in \Eq{eq:final_estimate}, and $\omega $ is chosen as three different values in range [1,1.5].
    The fits are performed using data for system sizes ranging from $L_\text{min}$ to $L_\text{max}=512$.
    }
    \label{tab:nline_mres}
\end{table}

Along the Nishimori line, we also compute the magnetic order parameter at the center of the system $m_{\text{center}}=\overline{m(L/2,L/2)}$. 
Near the MNP, it is expected to obey the finite-size scaling form
\eq{
m_\text{center} = L^{-\Delta} \left(f((p-p_c)L^{1/\nu})  + L^{-\omega} g((p-p_c)L^{1/\nu})\right),
}{eq:m_scaling}
where $\Delta$ is the scaling dimension of the order parameter.
Table~\ref{tab:nline_mres} reports the fits of $m_{\text{center}}$ for various $L_\text{min}$ values, with $p_c$, $1/\nu$ fixed to the values in \Eq{eq:final_estimate}.
To stabilize the fits, the correction exponent $\omega$ is also fixed within the range $[1,1.5]$ (which is reasonable from the previous fitting).
We find that the results are relatively insensitive to the fixed values of $p_c$, $1/\nu$ but show a noticeable dependence on the chosen $\omega$.
As $L_\text{min}$ increases, the resulting $\Delta$ have different trends for large and small $\omega$ and approach each other, and the quality of the fits also improves.
We omit the results of $L_\text{min}=8$, where the $\chi^2/\text{dof}$ is substantially larger than 1, and obtain our final estimate to contain all the fitting values with $L_\text{min}>8$:
\eq{
\Delta = 0.0893(3),\ \eta = 2\Delta =  0.1786(6).
}{}

\textit{Details of the simulation and finite-size scaling results away from the Nishimori line.}

\begin{table}[]
\setlength{\tabcolsep}{10pt}\centering
    \begin{tabular}{c|c|c|c|c}\hline
          $L$& $T_\text{min} -T_c$&$T_\text{max}-T_c$&  $\Delta T$ &$n_\text{sample}$\\\hline
          8-32& -0.14&0.2& 
    0.02&$6\times 10^7$\\\hline
 48,64& -0.1& 0.16&0.02&$6\times 10^7$\\\hline
 96,128& -0.06& 0.12& 0.02&$6\times 10^7$\\\hline
 256& -0.05& 0.75& 0.025 &$1.2\times 10^7$\\\hline
 512& -0.05& 0.05& 0.05&$1.2\times 10^7$\\ \hline\end{tabular}
    \caption{Simulation parameters on the vertical line at $p_c$ in \Fig{fig:tline}.}
    \label{tab:tline_params1}
\end{table}

\begin{table}[]
 \setlength{\tabcolsep}{10pt}
    \centering\begin{tabular}{c|c|c|c|c}\hline
          $L$& $(T_\text{min} -T_c)L^{1/4}$&$(T_\text{max}-T_c)L^{1/4}$&  $\Delta TL^{1/4}$&$n_\text{sample}$\\\hline
          8-128& -0.04&0.04& 
    0.002&$6\times 10^7$\\\hline
 256& -0.04& 0.04&0.002&$1.2\times 10^7$\\ \hline\end{tabular}
    \caption{Simulation parameters on the vertical line at $p_c$ used for finite-size scaling analysis. Note that $T=T_c$ points are not calculated twice, as they are already included in Table.~\ref{tab:tline_params1}.}
    \label{tab:tline_params2}
\end{table}
Table~\ref{tab:tline_params1} lists the simulation parameters used in Fig.~\ref{fig:tline}, while Table~\ref{tab:tline_params2} lists those used to conduct the finite-size scaling analysis below.
These simulations are performed at fixed $p=p_c$ , corresponding to the vertical line in the phase diagram, which represents the second relevant direction at the MNP.
Here, we actually take $p_c=0.1092211$ from an earlier estimation,
which still lies inside our reported uncertainty range for $p_c$ and has a negligible effect on the following fitting results.
The temperature window used in Table~\ref{tab:tline_params1} is much broader than that in Table~\ref{tab:tline_params2}, allowing the extreme behavior of various quantities to be clearly demonstrated.
However, because of this large window size and the small value of $1/\nu_T$, the data in Table~\ref{tab:tline_params1} do not provide reliable estimates of the critical exponents.
Therefore, in Table~\ref{tab:tline_params2} we adopt a much narrower temperature window and apply a $L^{-1/4}$ scaling so that the rescaled $(T-T_c)L^{1/\nu_T}$ remains approximately the same for different system $L$.

\begin{table}[]
    \centering
    \begin{tabular}{c|c|c|c|c|c|c|c|c|c}\hline
          &$L_\text{min}$&  $\chi^2/\text{dof}$ & $1/\nu_T$&$\omega$ &$a_0$& $a_1$& $a_2$&$b_0$ &$b_1$ \\\hline
 ${\mathcal{P}_\text{succ}^\text{MLD}}^*$&8 &222.09/182 &0.2506(10) & 1.179(14) & 0.8403555(48) & -0.15762(71) & -0.0964(32) & 0.01482(41) & 0.0040(57)\\\hline
 & 16& 173.43/161 &0.2518(16) & 1.197(36) & 0.8403615(65) & -0.1567(12) & -0.1011(34) & 0.0155(15) & -0.012(17)\\\hline
& 24 &151.18/140 &0.2515(23) & 1.192(75) & 0.8403629(93) & -0.1569(18) & -0.1047(39) & 0.0153(34) & -0.007(33)\\\hline
& 32 &132.30/119 &0.2507(32) & 1.12(12) & 0.840358(14) & -0.1576(26) & -0.1087(48) & 0.0119(46) & 0.006(47)\\\hline
 $d_W$& 8& 286.98/182 &0.2492(10) & 1.2940(55) & 2.241234(75) & -2.524(11) & 0.554(55) & -0.8515(94) & 0.85(12)\\\hline
 & 16& 190.59/161 &0.2503(14) & 1.374(14) & 2.240914(86) & -2.511(16) & 0.455(50) & -1.061(40) & 0.73(38)\\\hline
  & 24&161.02/140 &0.2501(20) & 1.390(31) & 2.24090(12) & -2.513(24) & 0.396(54) & -1.11(10) & 0.88(84)\\\hline
 & 32& 133.32/119 &0.2507(26) & 1.433(54) & 2.24084(14) & -2.506(32) & 0.335(58) & -1.29(22) & 0.6(16)\\\hline
  $\overline{\mathcal R^{0.1}}$& 8&252.62/182 &0.2493(12) & 1.3083(77) & 0.8202572(56) & 0.16166(84) & 0.0082(41) & 0.04762(73) & -0.0232(88)\\\hline
& 16& 183.55/161 &0.2502(17) & 1.396(21) & 0.8202749(66) & 0.1609(13) & 0.0144(39) & 0.0606(34) & -0.008(31)\\\hline 
& 24& 159.64/140 &0.2499(25) & 1.397(46) & 0.8202731(92) & 0.1612(19) & 0.0178(43) & 0.0608(84) & -0.018(67)\\\hline 
& 32& 134.05/119 &0.2508(32) & 1.461(81) & 0.820278(11) & 0.1604(25) & 0.0219(46) & 0.075(20) & 0.03(14)\\\hline 
 $1$$-$$\mathcal S_\text{DW}$& 8& 249.98/182 &0.24957(78) & 1.75(15) & 0.4989822(85) & -0.4218(14) & -0.1676(77) & -0.0146(46) & -0.206(64)\\\hline 
& 16& 189.06/161 &0.25000(92) & 2.81(69) & 0.4989825(76) & -0.4210(17) & -0.1815(77) & -0.29(55) & -6(11)\\\hline 
& 24& 158.51/140 &0.2500(11) & 4.1(31) & 0.4989879(76) & -0.4211(20) & -0.1930(82) & -19(192) & -423(4075)\\\hline 
&32 &134.52/119 &0.2505(14) & 12(186) & 0.4989926(75) & -0.4201(26) & -0.2005(90) & --- & ---\\\hline 
combined
& 32 &--- &0.2511(23) & 1.410(44)&---&---&---&---&---\\\hline

 \end{tabular}
    \caption{
    {\bf Temperature perturbation of MNP}.    
    Fitting results using different quantities along the vertical line at $p=p_c$. Here, ${\mathcal{P}_\text{succ}^\text{MLD}}^*$ is not the actual success probability of MLD; it uses the formula ${\mathcal{P}_\text{succ}^\text{MLD}}^*=\overline{\frac{\text{max}(\mathcal Z,\mathcal Z')}{\mathcal Z+\mathcal Z'}}$ which coincides with $\mathcal{P}_\text{succ}^\text{MLD}$ only along the Nishimori line.
    The last row gives the combined fitting results using all the above quantities with shared $1/\nu_T, \omega$ and bootstrap method.
    }
    \label{tab:tline_res}
\end{table}

\begin{figure*}
    \centering
    \includegraphics[width=1.0\linewidth]{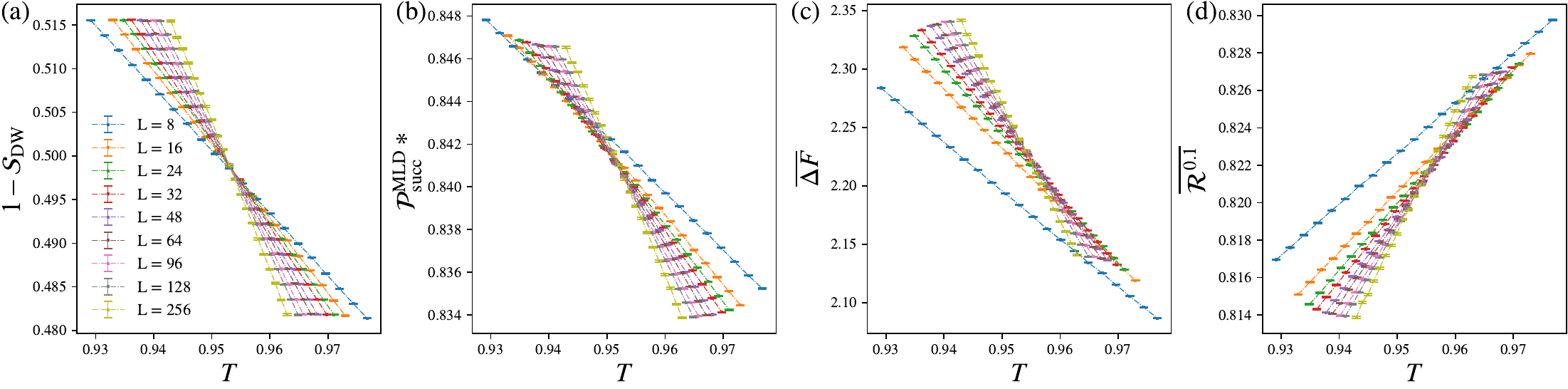}
    \caption{\textbf{Behavior of quantities under small perturbations away from the Nishimori line at $p = p_c$.} 
    (a) $1 - \mathcal{S}_\text{DW}$; 
    (b) ${\mathcal{P}_\text{succ}^{\text{MLD}}}^*$, defined as $\overline{\frac{\max(\mathcal{Z}, \mathcal{Z}')}{\mathcal{Z} + \mathcal{Z}'}}$; 
    (c) domain-wall free energy (DWFE) average $\overline{\Delta F}$; 
    (d) $\overline{\mathcal{R}^{0.1}}$. 
    The data shown correspond to the data used to produce the fitting results in Table~\ref{tab:tline_res}.
    }
    \label{fig:S4}
\end{figure*}

Using the data from simulations in Table.~\ref{tab:tline_params2}, we fit various quantities using the same finite-size scaling form as in \Eq{eq:fss}, but with $n_f= 3, n_g =2$ ($n_g=1$ will give larger $\chi^2/\text{dof}$ for several observables). 
Along the vertical line at MNP, the correlation length is expected to be governed by a distinct critical exponent $\nu_T$.
Table.~\ref{tab:tline_res} summarizes the fitting results for different quantities.
We do not include quantities like $\overline{\mathcal R^{0.5}}, \mathcal I_c(\beta)$, that exhibit extreme behavior along the Nishimori line, as discussed in the main text.
We also do not include $\mathcal{P}_\text{succ}^\text{Bayes}, \overline{\mathcal{R}^{0.3}}$ in the analysis, as their statistical uncertainties are significantly larger than those of the observables listed in Table~\ref{tab:tline_res} (as no variance reduction estimator can be used for $\beta \neq \beta_p$). 
As shown in the table, the quality of fit, measured by $\chi^2/\text{dof}$, approaches unity for $L_\text{min} \geq 16$. Among all observables studied, the domain-wall entropy exhibits the weakest finite-size effects (see comparison of these quantities in \Fig{fig:S4}). 
For $L_\text{min}=32$, the fitting of $b_0,b_1$ becomes even unstable.
Similar to the analysis along the Nishimori line, we perform a combined fit using all observables listed in Table~\ref{tab:tline_res}. 
A bootstrap procedure with 50,000 samples is again employed to account for correlations among the different quantities, yielding $1/\nu=0.2511(23)$, as reported in the last row of the table.
This combined fit results in a larger uncertainty than that obtained from the fit using $1-\mathcal{S}_{DW}$. 
Nevertheless, we adopt the combined result as our final estimate, since it consistently includes all individual fitting values, leading to
\eq{
1/\nu_T = 0.251(2).
}{}
\end{document}